\documentclass{pasj00}

\begin{document}
\SetRunningHead{S. Watanabe and H. Shibahashi}
{Seismic Solar Models and the Solar Neutrino Problem}
\Received{2000/12/28}
\Accepted{2001/04/25}

\title{Solar Models with Helioseismic Constraints\\
and the Solar Neutrino Problem}

\author{Satoru \textsc{Watanabe} and Hiromoto \textsc{Shibahashi}}
\affil{Department of Astronomy, School of Science, 
The University of Tokyo, 
Bunkyo-ku, Tokyo 113-0033 }
\email{watanabe@astron.s.u-tokyo.ac.jp}


%


\maketitle

\begin{abstract}
Imposing a constraint of the sound-speed profile 
determined from helioseismology and updating the microphysics, we 
have revised 
our seismic solar model, constructed with the assumption of a 
homogeneous 
metal abundance distribution, and have shown that the theoretically 
expected 
neutrino fluxes are still significantly more than the observations. 
With the same sound-speed profile constraint, we also constructed solar 
models with low metal abundance in the core, and evaluated the neutrino 
fluxes 
of these models to see if nonstandard solar models with a low metal 
core can 
solve the solar neutrino problem. Some of these models are in agreement 
with 
the Homestake data, the Super-Kamiokande data, and the sound-speed 
profile 
simultaneously, but none of these satisfy both the neutrino flux data, 
including GALLEX and SAGE, and the helioseismically determined density 
profile.
\end{abstract}

\section{Introduction}
Various nonstandard solar models have been proposed to explain the 
deficit of 
neutrino fluxes from the Sun compared with the theoretical 
expectation 
based on the standard evolutionary models of the Sun. Solar models 
having a 
core with low heavy element abundance, $Z$, are such examples 
(Joss 1974; Levy,
Ruzmaikina 1994; Jeffery et al.\ 1997). It is expected 
that a low $Z$ in the central region of 
the Sun would lead to low opacity there, resulting in a lower central 
temperature; such models are expected to be in better agreement with 
the 
observed neutrino fluxes. However, nonstandard solar models have often 
been 
criticized: either their properties are in substantial disagreement with
seismic data, or their sound-speed profile disagrees significantly with 
the 
seismically determined sound-speed profile of the Sun (e.g. Bahcall 
et al.\ 1997; Gough 1999). For example, comparing the observed data and 
the model
predictions of the frequency difference between the $n$-th overtone 
p-mode 
of the low-degree $l$ and the $(n-1)$-th overtone p-mode of the degree 
$l+2$, 
which is sensitive to the structure of the solar central region, 
Guenther and 
Demarque (1997) claimed that low-$Z$ cores extending beyond 
$0.06 M_{\odot }$ 
should be ruled out because, otherwise, the model predictions would be 
in
disagreement with the observed data and, furthermore, there would be 
no chance that the 
low-$Z$ core models could reduce the neutrino capture rates below 
$\sim $4.87 
SNU and $\sim $119 SNU for the Homestake neutrino detector using 
$^{37}$Cl and 
the GALLEX and SAGE detectors using $^{137}$Ga, respectively, while the 
detected capture rates are $\sim $2.56 $\pm$ 0.23 SNU and $\sim $72 
$\pm$ 8 
SNU, respectively (A SNU is defined to be $10^{-36}$ 
interactions s$^{-1}$ per target 
atom).
Such criticisms are, however, not necessarily fair, because those 
models 
were constructed without fine tuning to 
agree with the helioseismic data.

The recent observations of solar oscillations provide us with a large 
number 
of very accurate eigenfrequencies of the Sun. The relative errors 
in the frequency measurement are as small as $10^{-6}$. With the 
inversion of 
these frequencies, we now know the sound-speed profile in the Sun, 
$c(r)$, 
within errors of a few tenths of a percent. Imposing the
constraint of the thus-obtained sound-speed profile, the profiles of 
temperature, density, pressure, luminosity, and chemical composition 
can be 
deduced (Shibahashi 1993; Shibahashi, Takata 1996).

It should be stressed here that the sound-speed profile has been very 
precisely determined from helioseismology, and that the difference 
between the 
profile determined in this way and that of any standard evolutionary 
solar model is demonstrably
larger than the observational error. 
This means that we can also determine the profiles of thermal 
quantities in the Sun by using the observationally based highly
accurate 
profiles for the acoustic quantities and adopting the same information 
of
the microphysics (equation of state, nuclear reaction rates, and 
opacity) 
as used in constructing standard evolutionary solar models. 
The benefits of building such a 
solar model are as follows (Shibahashi 1993, 1999; 
Takata, Shibahashi 1998a, hereafter TS98a):
\begin{itemize}
\item We can construct a model of the present-day Sun without any 
assumption about the evolutionary history of the Sun. For example, 
we need not worry whether the $^3$He 
induced g-mode instability and/or mass loss affects the evolution of 
the Sun.
\item The location of the base of the convection zone, $r_{\rm conv}$, 
can be
estimated from the fact that the sound-speed profile changes rapidly 
there
(Gough 1984). Knowing the depth of the convection zone allows us to 
treat 
the radiative core only without worrying about the treatment of 
convection, 
which is not well-described theoretically 
(TS98a). 
\item The model is naturally consistent with the helioseismically 
determined sound-speed profile, while the evolutionary solar models 
are not necessarily so.
\end{itemize}

The main aim of this paper is to discuss our construction of
solar models with an assumption of low $Z$ in the core 
with the imposition of a constraint of the sound-speed profile, 
and to evaluate the neutrino fluxes of these models to see if 
nonstandard 
solar models with a low-$Z$ core can solve the solar neutrino problem.
Solar models based on helioseismology were constructed with an
assumption that $Z$ is homogeneous throughout the Sun (TS98a).
Since then, the nuclear cross sections have been systematically 
recompiled and
updated. Before discussing low-$Z$ core models of the Sun, 
in section 3 we revise the seismic solar model with a homogeneous 
$Z$ by adopting updated input physics. Then, in section 4,
with the same input physics, we construct seismic solar models having a 
low-$Z$
core and evaluate their neutrino fluxes.
The preliminary results were presented at the SOHO-10/GONG-2000 joint 
workshop
(Watanabe, Shibahashi 2000). We describe here our method in more 
detail, and
present the latest results, including an estimate of the errors in the
model properties and in the neutrino fluxes.

\section{Equations of the Seismic Solar Model}
We solve the basic equations governing the radiative core of the Sun 
with the 
imposition of information from helioseismology (TS98a, 
Takata, Shibahashi 1998b, hereafter TS98b, 1999; 
Shibahashi et al.\ 2000). They are 
the continuity equation,
\begin{equation}
	\frac{dM_r}{dr} = 4 \pi r^2 \rho,
\label{eq:1}
\end{equation}
the hydrostatic equation,
\begin{equation}
	\frac{dP}{dr} = -\frac{G M_r \rho}{r^2},
\label{eq:2}
\end{equation}
the energy equation,
\begin{equation}
	\frac{dL_r}{dr} = 4 \pi r^2 \rho \varepsilon,
\label{eq:3}
\end{equation}
and the energy transfer equation,
\begin{equation}
	\frac{dT}{dr} 
	= 
	-{{3}\over{4ac}} {{\kappa\rho}\over{T^3}} 
	{{L_r}\over{4\pi r^2}},
\label{eq:4}
\end{equation}
where the symbols have their usual meanings.

In addition to these differential equations, we need auxiliary 
equations: 
the equation of state, the equations for the opacity and the nuclear 
reaction rates:
\begin{equation}
	\rho = \rho (P, T, X_i),
\label{eq:5}
\end{equation}
\begin{equation}
	\kappa = \kappa (P, T, X_i),
\label{eq:6}
\end{equation}
and
\begin{equation}
	\varepsilon = \varepsilon (P, T, X_i).
\label{eq:7}
\end{equation} 
These equations link the thermal quantities and the chemical abundances.
In making standard evolutionary models, the chemical abundance 
distribution is 
obtained by following the time evolution,
\begin{equation}
	{{\partial X_i}\over{\partial t}} =  
	\left({{\partial X_i}\over{\partial t}}\right)_{\rm nuclear}	
	+ \left({{\partial X_i}\over{\partial t}}\right)_{\rm diffusion}
,
\label{eq:8}
\end{equation} 
where the nuclear reactions and the diffusion processes are taken into 
account.
In the standard scenario, the chemical composition is assumed to be 
uniform
at zero-age; $\partial X_i(t=0)/\partial r = 0$.  
In many nonstandard evolutionary models, some other time 
evolution processes are introduced and/or the initial conditions at 
$t=0$ are different from the standard case.

On the other hand, in constructing seismic models, we do not follow the 
time evolution. Instead, we impose the seismically determined 
sound-speed profile as a constraint on the model. Note that, 
if we distinguish only the hydrogen ($^1$H) and helium ($^4$He) 
separately 
as $X$ and $Y$, respectively, and treat all other elements 
collectively 
as heavy elements, $Z$, then the sound-speed, as one of the 
thermodynamical 
quantities, is a function of two other thermodynamical quantities 
($P$ and $T$) and $X$ and $Z$,
\begin{equation}
	c (P, T, X, Z) = c_{\rm obs}(r).
\label{eq:9}
\end{equation}
This inversely relates the hydrogen abundance, $X$, 
with the pressure, temperature, metal abundance, and the sound-speed, 
\begin{equation}
	X = X (P, T, Z, c_{\rm obs}).
\label{eq:10}
\end{equation}
Since $X$ is given by equation (\ref{eq:10}) with the helioseismically 
determined sound-speed, the density, the opacity, and the nuclear 
reaction rate are, in 
turn, given in terms of $(P, T, Z, c_{\rm obs})$ by equations (5)--(7).

If $Z$ is given, then with the help of equation (10), the basic
equations (1)--(4) can be solved with the proper boundary
conditions.
In principle, the   
helioseismically determined density profile, $\rho_{\rm obs}(r)$, can be 
used as another constraint in constructing 
a seismic solar model. 
This constraint together with the sound-speed constraint 
determines the heavy elements' abundance profile, $Z$, at a given 
$r$ as 
a part of the solution. Such attempts have indeed been carried 
out (TS98b). They have, however, not yet succeeded in obtaining a 
reasonable 
$Z$-profile in this way, since the dependence of the equation of state
upon $Z$ is so weak. We should also note that the formal error of 
$\rho _{\rm obs} (r)$ is much larger than that of $c_{\rm obs} (r)$.
In this paper, therefore, we impose only the sound-speed profile 
as a constraint in constructing a seismic solar model, and we 
assume a certain $Z$-profile a priori. 

In section 3, $Z$ is assumed to be homogeneous for simplicity, following
TS98a, while the microphysics is updated. In section 4, we assume that 
in the 
central core the heavy elements' abundance, $Z$, is lower than 
the outer part:
low-$Z$ core models. The schematic $Z$-profiles are shown in figure 1.
A typical $Z$-profile of a standard evolutionary model with 
gravitational settling is shown in figure 1 c (e.g., Bahcall et al.\ 
1998, hereafter we refer to 
it as BP98). 
TS98b treated a rectilinear $Z$-profile and their 
$Z$-profile is just like figure 1 d.

The boundary conditions at the center are trivial: 
\begin{equation}
	M_r=0 ~~~{\rm and}~~~ L_r=0 ~~~{\rm at}~~~r=0.
\label{eq:11}
\end{equation}
The location of the base of the convection zone is estimated from 
the inverted sound-speed profile (Gough 1984). 
We adopt the seismically determined depth of the convection zone,
$r = r_{\rm conv}$, and set 
the outer boundary conditions at the base of 
the convection zone (TS98a):
\begin{eqnarray}
	\nabla_{\rm ad} = \nabla_{\rm rad} &\equiv& 
	{{3\kappa L_r}\over{16\pi ac G M_r T^4}}
	~~~{\rm and}~~~ 
	L_r = L_{\odot} \nonumber\\
~~~&{\rm at}& ~~~r = r_{\rm conv}.
\label{eq:12}
\end{eqnarray}
This means that we do not need to care about the convective heat 
transport, 
which has theoretical uncertainties. 
Moreover, chemical homogeneity in the convection zone requires that
$Z/X$ at the base of the convection zone, $(Z/X)_{\rm conv}$, should be 
identical with the value at
the photosphere, $(Z/X)_{\rm surf}$, which is determined 
spectroscopically. 

Table 1 summarizes the microphysics and parameters of
the seismic solar models adopted in the present work. 
We adopt the updated microphysics which was also used by BP98.
Table 2 summarizes the differences in some important nuclear 
cross-section factors between TS98a and this work (Adelberger et al. 
1998). 
As for the helioseismically determined sound-speed profile, we adopt 
Basu's (1998) results obtained by inversion of the 
MDI velocity data of the first 144 days.
The formal error of the sound-speed inversion is smaller than $0.05$\% 
in the region $0.2 \le r/R_{\odot} \le 0.8$,
and does not exceed $0.3$\% in the entire region.
In the central core ($r/R_{\odot}\le 0.05$),
we extrapolate the inverted sound-speed profile and its error level to 
the 
center. 

\section{Structure of the Seismic Solar Model with Homogeneous $Z$}
We first construct a seismic solar model by assuming that $Z$ is 
homogeneous. The methodology is exactly the same as that adopted by 
TS98a.
The properties of the homogeneous $Z$ model are summarized in table 3. 

The effect of various uncertainties in microphysics upon the seismic 
model 
and the theoretically expected neutrino fluxes was investigated by 
a Monte Carlo simulation. For example, as for each of the nuclear cross 
sections,
we constructed 
one hundred sets of seismic models by superimposing Gaussian noise 
corresponding to the 1-$\sigma$ level uncertainty of each one of the 
astrophysical $S$-factors, while keeping other microphysics unchanged so
as to
isolate the effect of each microphysics. 
As for the inverted sound-speed profile, we also constructed one 
hundred sets 
of seismic solar models by adding Gaussian noise to the most likely 
sound-speed at every step of $\Delta r/R_\odot = 0.01$. In this process,
we 
interpolated the sound-speed at the other mesh points smoothly.
We estimated the effect of the uncertainties in opacity by constructing
an additional one hundred sets of seismic models with opacities having 
Gaussian 
noise with 5\% amplitude added in the whole range of density, 
temperature and 
chemical compositions at all mesh points. To see the effect of the 
equation of state, we constructed a seismic solar model with the 
ideal gas law, 
and compared the model structure with that constructed with the OPAL 
equation 
of state.
Table 4 summarizes the effect of the uncertainties in microphysics,
investigated in this way,  
upon the theoretically expected neutrino fluxes, the central 
density, $\rho _{\rm c}$, and 
the helium abundance at the surface and the base of
the convection zone, $Y_{\rm conv}$, of the seismic solar model.

It is clear from this table that the influences of the $S(0)$-factors 
of 
$^1$H(p,e$^+\nu_{\rm e}$)$^2$H-, $^3$He($^4$He,$\gamma$)$^7$Be-, and 
$^7$Be(p,$\gamma$)$^8$B-reactions are crucial for the neutrino fluxes. 
The increase in the $S_{11}(0)$-factor makes the nuclear energy
generation more efficient. Since the luminosity should be fixed, this
means that the temperature and density near the center should become
lower. Since the sound-speed ($c^2 = \Gamma _1 P/\rho \approx
\Gamma _1 RT/\mu $) is also fixed, the decrease in the
temperature, $T_{\rm c}$, and in the density, $\rho _{\rm c}$, leads to
an increase in the hydrogen abundance, $X_{\rm c}$, and a decrease 
in the pressure, $P_{\rm c}$, respectively.
Decreases in $\rho_{\rm c}$ and $T_{\rm c}$ make the pp-II and the 
pp-III
reactions less efficient. Thus, the $^7 $Be- and the $^8$B-neutrino
fluxes decrease and, hence, the neutrino capture rates for both the
chlorine experiment and Super-Kamiokande become smaller. The decrease in
the $^7 $Be- and the $^8$B-neutrino fluxes dominates over a slight
increase in the pp-neutrino flux; the neutrino capture rate for the
gallium experiments also decrease. The increase in the
$S_{33}(0)$-factor makes the pp-I reaction more efficient, and thus the
contributions of the pp-II- and pp-III reactions to the energy 
generation
become smaller. Hence, the $^7 $Be- and the $^8$B-neutrino fluxes 
caused 
by the pp-II and the pp-III reactions, respectively, decrease, and then
the neutrino capture rates for the chlorine experiment and the
Super-Kamiokande decrease.
Decreases in the $^7 $Be- and the $^8$B-neutrino fluxes also lead
to a decrease in the neutrino capture rate for the gallium
experiments. 
On the other hand, an increase in the $S_{34}(0)$-factor makes the
branching ratio of the pp-II and pp-III reaction to the pp-I reaction
larger. Consequently, the $^7 $Be- and the $^8$B-neutrino fluxes
increase, and the neutrino capture rates for all of the neutrino 
detection experiments increase. 
The increase in the  $S_{17}(0)$-factor makes the branching ratio of 
the 
pp-III reaction to the pp-II reaction larger. Hence it makes the
$^8$B-neutrino flux and the neutrino capture rates for the chlorine
experiment and the Super-Kamiokande particularly larger. 
For the neutrino capture rate for the gallium experiments, the 
uncertainty 
in the neutrino cross-section is also crucial because the uncertainty 
is larger than
that for the chlorine detector. 
On the other hand, the mass ratio of metal to hydrogen at the surface, 
$(Z/X)_{\rm surf}$, the depth of the convection zone, 
$r_{\rm conv} /R_{\odot}$, the opacity, and the equation of state are 
crucial 
for the surface helium abundance, $Y_{\rm conv}$. Among them, 
$(Z/X)_{\rm surf}$ is crucial because it is related 
to $Y_{\rm conv}$ directly and
its uncertainty is large.

A comparison of the present result with observations and some other 
solar models are summarized in table 5 and depicted in figure 
2. Note that errors of TS98a and TS98b are rough estimations. 
The seismic solar model presented by TS98a was constructed with the
same assumption of $Z$ as in the present case. We thus regard the
present model as a revised version of TS98a. While the sound-speed 
profile is replaced with the updated one,
the difference is so small that this replacement does not induce 
a significant 
difference in the structure of the model. Rather, the difference in the 
model structure comes mainly from the updated input microphysics.
As can be seen in table 5, 
the neutrino-capture rates actually detected in the current experiments
are still 
significantly less than predictions based on the present seismic 
solar 
model.

Figure 3 shows the relative differences in the sound-speed profile 
and in the density profile between the present seismic solar model and 
a 
direct inversion of the oscillation data (Basu 1998). 
For a comparison, the differences between the standard evolutionary 
model (BP98) and the direct inversion are also shown.
Since we imposed a constraint of the inverted sound-speed profile in 
constructing the seismic solar model, the relative difference in the 
sound-speed of the seismic model shown in the left panel is exactly 
equal to 
the amount of the uncertainty in the inverted sound-speed
profile, itself. 
Note that the difference between the sound-speed profile of the 
evolutionary 
standard model (e.g., BP98) and the helioseismically inverted profile 
is 
larger than the 1-$\sigma$ level of uncertainty. The seismic model is 
consistent with the inverted density profile only within the 
2-$\sigma $ level, 
while the difference between the density profile of the evolutionary 
standard
model (BP98) and the inverted profile is larger than the 2-$\sigma $ 
level near the base of the convective zone. 

The uncertainty in the density profile of the seismic solar model 
is, however, larger than the 1-$\sigma $ level uncertainty of the 
inverted 
result. Note that the origin of each uncertainty is different. The 
uncertainty of the inverted result is caused by
the measurement
errors of solar oscillations. In order to see what is the main cause of
the large uncertainty of the seismic solar model, we carried out a
Monte Carlo simulation while changing only one input
microphysics, and keeping all others unchanged to isolate the effect 
of each microphysics. Figure 4 shows the uncertainty in the density
profile, caused only by each input microphysics. 
The influence of the highly accurate sound-speed profile is very small.
From this simulation, it is concluded that the main causes are
the 
nuclear cross-sections of the pp-reaction [$S_{11} (0)$] and of the 
$^3$He+$^4$He-reaction [$S_{34} (0)$]. Since energy generation in
the Sun is mainly controlled by $S_{11} (0)$, it is most
crucial for the solar structure: density profile. Although $S_{34} (0)$
is less important than $S_{11} (0)$ for the solar structure, its
uncertainty is much larger. 

Figure 3b shows that the uncertainty in the density profile of the
seismic model is very small around $r = 0.2R_{\odot }$. Since the
total luminosity is fixed, with the increase of $S_{11} (0)$, which is
most important for the density profile, the density in the core 
decreases to compensate for the increase in the nuclear reaction rate. 
On the other
hand, since the total mass is fixed, the density outside the core
increases to compensate for the decrease of the density in the core. 
Because 
these opposite reactions of the density against the input physical
parameter balance around $r = 0.2R_{\odot }$, the uncertainty in the
density of the seismic model is very small around there. 

\section{Structure of the Seismic Solar Model with a Low-$Z$ Core}
Solar models having a low-$Z$ core are expected to lead to low neutrino
fluxes. One possible scenario for the formation of a low-$Z$ core is to
assume that dust and heavy elements remaining in the disk around the
Sun after the formation of the planets were accreted onto the Sun so 
that heavy elements would be accumulated more in the envelope than 
in the core
(Joss 1974; Jeffery et al.\ 1997). In another
unconventional scenario, heavy elements locked up in grains are assumed
to be segregated from the hydrogen and helium gas in the pre-solar 
nebula 
(Levy, Ruzmaikina 1994). In
order to see how much the neutrino fluxes can be reduced by 
introducing a
low-$Z$ core, we constructed seismic solar models having a low-$Z$ core 
with the
imposition of the sound-speed profile.
We assume here that the $Z$-profile is a step function of $r$; that 
is, $Z(r)$ is a certain constant, $Z_{\rm c}$, 
in the region $0 \leq r \leq r_{\rm f}$, where $r_{\rm f}$ is a 
parameter, while $Z$ is another constant in the region 
$r_{\rm f} \leq r \leq r_{\rm conv}$, so that $Z/X$ matches its surface 
value,
$(Z/X)_{\rm surf}$ (figure 1 b).
The parameter $Z_{\rm c}$ ranges from $0.0001$ to $0.020$.
We constructed the models so that their sound-speed profiles would be 
consistent with the seismically determined profile. Table 6 shows the 
properties of low-$Z$ core models with various values of 
$r_{\rm f}$ in 
the case of $Z_{\rm c}=0.0001$.
Note that the model with $r_{\rm f}=0$ is the same as the homogeneous
$Z$ model discussed in the previous section. In this table, 
$M_{\rm core}$ denotes the mass of the low-$Z$ core 
[$M_{\rm core} \equiv M_r (r = r_{\rm f})]$.

The properties of the low-$Z$ core models can be explained as follows. 
With an increase in the low-$Z$ core size, the opacity in the core 
decreases,
resulting in a decrease of the central temperature. Since 
the sound-speed
profile is fixed by the observation, this requires a decrease in the 
mean molecular weight to
compensate for a decrease in the temperature, and hence an increase in 
the
hydrogen abundance there. The total luminosity is also fixed, which 
requires an
increase in the density in the core to compensate for the decrease 
in the
nuclear reaction rate due to the decrease in temperature there
[cf. equation (3)]. The
constraint of the sound-speed leads to an increase in the pressure 
in the
core because of the higher density. These tendencies are common in the
seismic solar models with a low-$Z$ core, and become conspicuous with 
a decrease in the value of $Z_{\rm c}$.

Figures 5 a-c show the dependence of the neutrino capture rates
of the seismic models upon the values of $Z_{\rm c}$ and $M_{\rm core}$.
For a
comparison, the detected neutrino capture rates are also indicated. The
neutrino capture rates for the chlorine detector (Homestake) and the 
$^8$B-neutrino flux (Super-Kamiokande) decrease substantially with an 
increase
in the low-$Z$ core size. 

The effect of various uncertainties in microphysics upon the model 
properties
and the theoretically expected neutrino fluxes was investigated by 
a Monte Carlo simulation in the same way as that described in the 
previous section.
The results are summarized in table 7 for the case of 
$Z_{\rm c}=0.0001$; also, 
the thus-determined error bars for the neutrino capture rates, 
the $^8$B-neutrino flux and the central density are depicted in 
figure 5.

Let us focus attention on those models with $Z_{\rm c} = 0.0001$
(the continuous line).
As can be seen in figure 5 a, the models with 
$M_{\rm core} > 0.1 M_\odot$ 
do not contradict the result of the Homestake experiment if we 
tolerate
the difference within the $2$-$\sigma$ level. In the same sense, 
the Super-Kamiokande experiment is consistent with those models 
having $0.05 M_{\odot} \leq M_{\rm core} \leq 0.12 M_{\odot}$.
That is, low-$Z$ core models with $M_{\rm core} \sim 0.1 M_\odot$ 
are
favorable for explaining the Homestake and the Super-Kamiokande 
experiments 
simultaneously. Besides that, such models are naturally consistent with 
the
seismically determined sound-speed profile. These results are not in 
agreement
with the conclusion of Guenther and Demarque (1997, hereafter we refer
to it as GD97), who investigated the 
low-$Z$ core model of the Sun by following its evolution. 
There are differences between the present approach and GD97's in the
concept and  methodology of making solar models.
While the models of GD97 are constructed following the stellar
evolution, the present models are not.
In this sense the present approach has more freedom.
We do not try to explain how our seismic models can be realized in the
evolutionary process, and such an investigation is beyond our scope. 
While the present models were constructed with fine tuning to agree with
the helioseismic data, those of GD97 were not. 
It also should be noted that GD97 adopted almost the 
same input physical parameters with TS98a. 
Updating them in GD97's approach may lead to
lower neutrino fluxes (cf. figure 2 and table 5), and the difference
between the present result and GD97's would become smaller.
Our results demonstrate that certain low-$Z$ core models of the Sun can 
be in
agreement with the Homestake and the Super-Kamiokande data and 
helioseismology.
Therefore, criticisms based on
evolutionary models are not necessarily fair. 
However, even the present seismic solar models are substantially 
in disagreement with the neutrino capture rates for the gallium 
detectors,
GALLEX and SAGE.  

In the present approach, we did not use the seismically obtained 
density profile as a constraint in making seismic solar models. We 
should compare the density profile of the models with the seismically 
inverted profile. In the case of $Z_{\rm c}=0.0001$, as shown in figure 
5 d, the central density of the
model becomes higher with an increase in the core size, 
and only those models with $M_{\rm core} < 0.02 M_\odot$
are consistent with the seismically inverted density (Basu 1998).
To see the effect of parameter $Z_{\rm c}$, we adopted various values 
of 
$Z_{\rm c}$ and constructed seismic models of the 
Sun with a low-$Z$ core in a similar way. However, no model is 
consistent with all of the neutrino experiments and the inverted 
density
profile. From these results, we conclude that none of the low-$Z$ core 
models 
are simultaneously consistent with all of the neutrino experiments and 
the helioseismically inverted density profile.

In the case of $Z_{\rm c} > Z_{\rm surf} (\approx 0.018)$, the 
temperature at 
the central region becomes higher than that of the homogeneous $Z$ 
model,
and hence deviations from the detected solar neutrino fluxes become 
larger, while the density at the central region becomes less 
inconsistent with the seismically inverted profile. 

\section{Summary and Conclusion}
We have revised the seismic solar model with a constraint of the 
sound-speed profile, and the latest input microphysics and physical 
parameters,
assuming that $Z$ is uniform, 
and also have estimated the uncertainty of the seismic model 
thoroughly by a
Monte Carlo simulation. Though the seismic solar model thus constructed 
is 
naturally consistent with the helioseismic data, the expected neutrino 
fluxes
based on this model are still higher than the experimental data.
This result implies that a modification of the solar model does not 
seem to be
able to solve the solar neutrino problem, leaving the 
neutrino-oscillation explanation as an attractive possibility. 
The uncertainty in the seismically determined sound-speed profile is so 
small that it does not induce substantial uncertainties in the physical 
properties of the model. 
The uncertainties in the nuclear cross-section [especially 
$S_{\rm 11}(0)$, $S_{\rm 34}(0)$, and $S_{\rm 17}(0)$] have more 
crucial influences. 

We have also constructed seismic models with an assumption of a
low-$Z$ core, and evaluated the neutrino fluxes of these models to see 
if these 
nonstandard solar models can solve the solar neutrino problem.
Although it turns out to be possible to explain the neutrino fluxes for 
both the Homestake and Super-Kamiokande experiments and the sound-speed 
profile 
simultaneously by the low-$Z$ core extending 
$\sim 0.01M_{\rm core}/M_{\odot }$, 
it is still impossible to explain all of the data, including the 
neutrino flux 
measurements based on the gallium detectors, by this model. 
The density in the core becomes higher with a decrease of 
$Z_{\rm c}$,
and the model becomes less consistent with a seismically inverted 
density 
profile.

Therefore, we conclude that the 
low-$Z$ core models of the Sun cannot explain all of the solar neutrino 
detection experiments and the helioseismically inverted sound-speed and
density profiles. \par
\vspace{1pc}\par
We would like to thank M. Takata for many useful discussions. One of 
the 
authors (SW) would like to express his sincere thanks to Professor H. 
Saio 
for guidance, continuous encouragement and many useful discussions 
during 
his fruitful campus life in Tohoku University. 
This research was supported in part by a Grant-in-Aid for Scientific 
Research
on Priority Areas by the Ministry of
Education, Culture, Sports, Science and Technology (12047208).

\onecolumn
\begin{table}
 \caption{The Microphysics and physical parameters of 
the seismic solar models.}
 \begin{center}
  \begin{tabular}{lll}
   \hline\hline
Physics/Parameter     &               & Reference \\
\hline
Equation of state     & OPAL          & Rogers et al.\ 1996\\
Opacity               & OPAL          & Iglesias, Rogers 1996\\
Nuclear cross-section &               & Adelberger et al.\ 1998\\
Screening effect      & Weak screening & Salpeter 1954; 
					 Gruzinov, Bahcall 1998\\
Neutrino cross-section &              & Bahcall 1997;
                                        Bahcall et al.\ 1996, 1998 \\
                      &	              & Garc\'\i a et al.\ 1991\\
$(Z/X)_{\rm surf}$      & $0.0245$      & Grevesse, Noels 1993\\

Luminosity $L_\odot$  & $3.844(1\pm 0.004)\times 10^{33} {\rm erg}$ 
s$^{-1}$  
& Bahcall, Pinsonneault 1995\\
\hline
Sound-speed profile
                      & SOHO data     & Basu 1998\\
$r_{\rm conv}$        & $(0.713\pm 0.001) R_{\odot}$
                      & Basu, Antia 1997\\
   \hline
  \end{tabular}
 \end{center}
\end{table}

\begin{table}
 \caption{Comparison of some important nuclear cross-section 
factors between TS98a and this work.}
 \begin{center}
  \begin{tabular}{ccccc}
   \hline\hline
  & \multicolumn{2}{c}{TS98a} & \multicolumn{2}{c}{This work} \\
Reference  & \multicolumn{2}{c}{Bahcall, Pinsonneault\ 1995} 
& \multicolumn{2}{c}{Adelberger et al.\ 1998} \\
\hline
  &  $S$(0) (keV barns) & $S^{'} (0)$ (barns) & $S$(0) (keV barns) 
& $S^{'}(0)$ (barns) \\
\hline
$^1$H(p,e$^+ \nu _{\rm e}$)$^2$H \dotfill 
& 3.89(1$\pm$0.011)$\times $10$^{-22}$ 
&  4.52$\times $10$^{-24}$ & 4.00(1$\pm 0.007^{+0.020}_{-0.011}$)
$\times $10$^{-22}$ 
& 4.48$\times $10$^{-24}$ \\
$^3$He($^3$He,2p)$^4$He \dotfill & 4.99(1$\pm$0.06)$\times $10$^{3}$ 
& $-0.9$ & (5.4$\pm$0.4)$\times $10$^{3}$ & $-4.1$ \\
$^3$He($^4$He,$\gamma $)$^7$Be \dotfill & 0.524(1$\pm$0.032) 
& $-3.1\times 10^{-4}$ 
& 0.53$\pm$0.05 & $-3.0\times 10^{-4}$ \\
$^7$Be(p,$\gamma $)$^8$B \dotfill & 0.0224(1$\pm$0.093) 
& $-3\times 10^{-5}$ 
& 0.019$^{+0.004}_{-0.002} (1\sigma ) ^{+0.008}_{-0.004} (3\sigma)$ 
& $-1.3\times 10^{-5}$\\
$^{14}$N(p,$\gamma $)$^{15}$O \dotfill & 3.29(1$\pm$0.12) 
& $-5.91\times 10^{-3}$ 
& 3.5$^{+0.4}_{-1.6} (1\sigma ) ^{+1.0}_{-2.0} (3\sigma )$ 
& $-1.28\times 10^{-2}$\\
   \hline
  \end{tabular}
 \end{center}
\end{table}

\begin{table}
 \caption{Properties of the seismic solar model with 
homogeneous
$Z$.}
 \begin{center}
  \begin{tabular}{cc}
   \hline\hline
Quantities & Most likely values \\
\hline
$T_{\rm c}$(10$^7$ K) \dotfill & 1.561 $^{+0.005}_{-0.009}$ \\
$P_{\rm c}$ (10$^{17}$ dyn cm$^{-2}$ ) \dotfill 
& 2.378 $^{+0.031}_{-0.049}$ \\
$\rho_{\rm c}$ (g cm$^{-3}$ ) \dotfill & 156.0 $^{+2.0}_{-3.3}$ \\
$X_{\rm c}$ \dotfill & 0.3383 $^{+0.0058}_{-0.0035}$ \\
$Y_{\rm c}$ \dotfill & 0.6437 $^{+0.0035}_{-0.0058}$ \\
\hline
$M_{\rm conv} /M_{\odot}$ \dotfill & 0.0264 $^{+0.0008}_{-0.0013}$ \\
$T_{\rm conv}$ (10$^6$ K) \dotfill & 2.19 $\pm 0.01$ \\
$P_{\rm conv}$ (10$^{13}$ dyn cm$^{-2}$ ) \dotfill 
& 5.71 $^{+0.07}_{-0.06}$ \\
$\rho_{\rm conv}$ (g cm$^{-3}$ ) \dotfill & 0.190 $\pm 0.001 $\\
$X_{\rm conv}$ \dotfill & 0.7365 $^{+0.0036}_{-0.0037}$ \\
$Y_{\rm conv}$ \dotfill & 0.2455 $^{+0.0042}_{-0.0041}$ \\
$Z_{\rm conv}$ \dotfill & 0.0180 $\pm 0.0004$ \\
\hline
pp $\nu$ flux (10$^{10}$ cm$^{-2}$ s$^{-1}$ ) \dotfill 
& 5.98 $^{+0.05}_{-0.04}$ \\
pep $\nu$ flux (10$^{8}$ cm$^{-2}$ s$^{-1}$ ) \dotfill 
& 1.44 $\pm 0.02$ \\
hep $\nu$ flux (10$^{3}$ cm$^{-2}$ s$^{-1}$ ) \dotfill 
& 2.11 $\pm 0.06$ \\
$^{7}$Be $\nu$ flux (10$^{9}$ cm$^{-2}$ s$^{-1}$ ) \dotfill 
& 4.72 $^{+0.39}_{-0.43}$ \\
$^{8}$B $\nu$ flux (10$^{6}$ cm$^{-2}$ s$^{-1}$ ) \dotfill 
& 4.77 $^{+1.04}_{-0.72}$ \\
$^{13}$N $\nu$ flux (10$^{8}$ cm$^{-2}$ s$^{-1}$ ) \dotfill 
& 4.43 $^{+0.49}_{-1.26}$ \\
$^{15}$O $\nu$ flux (10$^{8}$ cm$^{-2}$ s$^{-1}$ ) \dotfill 
& 4.15 $^{+0.49}_{-1.26}$ \\
$^{17}$F $\nu$ flux (10$^{6}$ cm$^{-2}$ s$^{-1}$ ) \dotfill 
& 5.22 $^{+1.08}_{-1.16}$ \\
$\nu$ capture rate for Cl (SNU) \dotfill & 7.17 $^{+1.24}_{-0.95}$ \\
$\nu$ capture rate for Ga (SNU) \dotfill & 126 $^{+6.6}_{-5.3}$ \\
   \hline
  \end{tabular}
 \end{center}
\end{table}

\begin{table}
 \caption{Sensitivity of the neutrino fluxes, the central 
density,
$\rho_{\rm c}$, and the surface helium abundance, $Y_{\rm conv}$, to 
the uncertainties in the input physics.$^*$}
 \begin{center}
  \begin{tabular}{rccccccc}
   \hline\hline
  &  &  & Cl    & Ga    & $^8$B & $\rho_{\rm c}$ & $Y_{\rm conv}$ \\
  &  &  & 7.17  & 126   & 4.77  & 156            & 0.246          \\
Input physics  
  &  &  & (SNU) & (SNU) & (10$^6$cm$^{-2}$s$^{-1}$) 
                                & (g cm$^{-3}$)  &                \\
\hline
pp           & 4.00(1$^{+0.021} _{-0.013}$)$\times$10$^{-22}$ 
                                 & (keV b) & $_{+0.372} ^{-0.633}$ 
                                                        &
$_{+1.8} ^{-3.1}$ 
           & $_{+0.282} ^{-0.477}$ 
                      & $_{+1.6} ^{-3.0}$ 
                                 & $^{+0.0008} _{-0.0003}$
                                               \\
pep          & $\pm$1\%          &         & $\pm$0.002 & 
0.0        & 0.000    & 0.0      & 0.0000      \\
$^3$He$^3$He & 5.4$\pm$0.32      & (MeV b) & $\mp$0.131 & 
$\mp$0.8 & $\mp$0.094 & $\pm$0.3 & $\mp$0.0001 \\
$^3$He$^4$He & 0.53$\pm$0.05     & (keV b) & $\pm$0.408 & 
$\pm$2.6 & $\pm$0.293 & $\mp$1.0 & $\pm$0.0003 \\
$^7$Be+e     & $\pm$2\% (1$\sigma$) 
                                 &         & $\mp$0.106 & 
$\mp$0.2 & $\mp$0.093 & 0.0      & 0.0000      \\
$^7$Be+p     & 19$^{+4} _{-2} (1\sigma ) ^{+8}_{-4} (3\sigma )$ 
                                 & (eV b)  & $^{+1.057} _{-0.456}$ 
                                                        & 
$^{+2.2} _{-1.0}$ 
         & $^{+0.927} _{-0.400}$ 
                      & 0.0      & 0.0000      \\
$^{12}$C+p   & 1.34$\pm$0.21     & (keV b) & 0.000      & 
0.0      & 0.000      & 0.0      & 0.0000      \\
$^{13}$C+p   & 7.6$\pm$1         & (keV b) & 0.000      &
 0.0     & 0.000      & 0.0      & 0.0000      \\
$^{14}$N+p   & 3.5
$^{+0.4} _{-1.6} (1\sigma ) ^{+1.0} _{-2.0} (3\sigma )$
                                 & (keV b) & $^{+0.023} _{-0.069}$ 
                                                        & 
$^{+0.5} _{-1.6}$ 
         & $_{+0.021} ^{-0.007}$ 
                      & $_{+0.6} ^{-0.2}$ 
                                 & $^{+0.0001} _{-0.0002}$ 
                                               \\
$^{16}$O+p   & 9.4$\pm$1.7       & (keV b) & $\pm$0.001 & 
0.0      & 0.000      & 0.0      & 0.0000      \\
$(Z/X)_{\rm surf}$ 
             & 0.0245$\pm$0.0006 &         & $\pm$0.120 & 
$\pm$0.5 & $\pm$0.094 & $\mp$0.2 & $\pm$0.0026 \\
$L_\odot$    & 3.844(1$\pm$0.004)$\times$10$^{33}$ 
                                 & (erg s$^{-1}$) 
                                           & $\pm$0.203 & 
$\pm$1.3 & $\pm$0.153 & $\pm$0.5 & $\pm$0.0005 \\
\multicolumn{3}{c}{Sound-speed profile}    & $\pm$0.030 & 
$\pm$0.1 & $\pm$0.022 & $\pm$0.3 & $\pm$0.0001 \\
$r_{\rm conv} /R_\odot$ 
             & 0.713$\pm$0.001   &         & $\mp$0.031 & 
$\mp$0.2 & $\mp$0.023 & 0.0      & $\pm$0.0031 \\
\multicolumn{3}{c}{Neutrino cross-section} 
                                           & $\pm$0.173 & 
$^{+5.1} _{-2.4}$ 
         & $\cdots$   & $\cdots$ & $\cdots$    \\
Opacity      & $\pm 5\%$         &         & $(\pm)$0.083 
                                                        &
$(\pm)$0.4 
         & $(\pm)$0.062 
                      & $(\pm)$0.2 
                                 & $(\pm)$0.0075 
                                               \\
EOS          & ideal ($\Gamma = \frac{5}{3}$) 
                                 &         & $-$0.007   & 
$-$0.3   & $+$0.006   & $-$0.6   & $+$0.012    \\
\hline
Total        &  $\sqrt{\sigma ^2 + \sigma ^2 + \cdots }$ 
                                 &         & $^{+1.24} _{-0.95}$ 
                                                        & 
$^{+6.6} _{-5.3}$ 
         & $^{+1.04} _{-0.72}$ 
                      & $^{+2.0} _{-3.2}$ 
                                 & $\pm$0.004  \\
\multicolumn{2}{c}{(except for opacity and EOS)}  
                                 &         & (SNU)      & 
(SNU)    & (10$^6$cm$^{-2}$s$^{-1}$) 
                      & (g cm$^{-3}$)
                                 &             \\
   \hline
  \end{tabular}
 \end{center}
$^*$ The first ten entries are due to the nuclear cross-section
factors. References for each uncertainty are the same 
as given in table 1 except for $(Z/X)_{\rm surf}$. The uncertainty of 
$(Z/X)_{\rm surf}$ is taken from Basu (1998).
\end{table}

\begin{table}
 \caption{Comparison of experiments and some solar models.}
 \begin{center}
  \begin{tabular}{cccccc}
   \hline\hline
 & Experiments & This work & TS98a & TS98b & BP98\\
\hline
 Cl(SNU) \dotfill & 2.56 $\pm $ 0.23 $^*$ 
& 7.17 $^{+1.24} _{-0.95} $
& 7.8 $\sim $ 10.6 & 7.7 $\sim$ 8.8 &  7.7 $^{+1.2} _{-1.0}$\\
\hline
 Ga(SNU) \dotfill & 67.2 $^{+8.0} _{-7.6}$ $^{\dagger}$ 
& 126 $^{+6.6} _{-5.3}$ 
& 129 $\sim$ 142  
& 132 $\sim$ 138
& 129 $^{+8} _{-6}$ \\
 &  77.5 $^{+7.5} _{-7.8} $ $^{\ddagger}$ & & & & \\
\hline
 $^8$B(10$^6$ cm$^{-2}$ s$^{-1}$) \dotfill & 2.80 $\pm$ 0.38 $^{\S}$ 
& 4.77 $^{+1.04} _{-0.72}$ 
& 5.48 $\sim$ 7.67
&5.3 $\sim$ 6.1
& 5.15 $^{+0.98} _{-0.72}$ \\
 & 2.42 $^{+0.12} _{-0.09}$ $^{\Vert}$ & & & & \\
\hline
\hline
$\rho_{\rm c} $(g cm$^{-3}$) \dotfill & 153.9 $\pm$ 1.1 $^{\sharp}$ 
& 156 $^{+2.0} _{-3.2}$ 
& 156 $\sim$ 171 
& 153 $\sim$ 155 
& $\sim $152.2 \\
\hline
$Y_{\rm conv}$ \dotfill & 0.248 $\pm$ 0.001 $^{\sharp}$ 
& 0.246 $\pm$ 0.004 
& 0.227 $\sim$ 0.236
& 0.246 $\sim$ 0.247
& $\sim$0.243 \\
   \hline
  \end{tabular}
 \end{center}
$^*$ Homestake : Cleveland et al.\ 1998
\par\noindent
$^{\dagger}$ SAGE : Abdurashitov et al.\ 1999
\par\noindent
$^{\ddagger}$ GALLEX : Hampel et al.\ 1999
\par\noindent
$^{\S}$ Kamiokande : Fukuda et al.\ 1996
\par\noindent
$^{\Vert}$ Super-Kamiokande : Fukuda et al.\ 1998
\par\noindent
$^{\sharp}$ Basu 1998
\end{table}

\begin{table}
 \caption{Properties of the low-$Z$ core model in the case 
of $Z_{\rm c}$ = 0.0001.$^*$}
 \begin{center}
  \begin{tabular}{ccccccccc}
   \hline\hline
$r_{\rm f}/R_\odot$ \dotfill & 0.000 & 0.050 & 0.069 & 0.083 & 0.10 
& 0.12 & 0.13 & 0.15 \\
$M_{\rm core}/M_{\odot}$ \dotfill & 0.000 & 0.013 & 0.030 & 0.050 
& 0.080 & 0.117 & 0.150 & 0.202 \\
\hline
$T_{\rm c}$(10$^7$ K) \dotfill & 1.56 & 1.53 & 1.51 & 1.49 & 1.48 
& 1.46 & 1.46 & 1.45 \\
$P_{\rm c}$ (10$^{17}$ dyn cm$^{-2}$ ) \dotfill & 2.38 & 2.40 & 2.41 
& 2.43 & 2.44 & 2.46 & 2.48 & 2.50 \\
$\rho_{\rm c}$ (g cm$^{-3}$ ) \dotfill & 156 & 157 & 158 & 159 & 160 
& 162 & 163 & 164\\
$X_{\rm c}$ \dotfill & 0.338 & 0.356 & 0.369 & 0.378 & 0.387 & 0.395 
& 0.399 & 0.403\\
$Y_{\rm c}$ \dotfill & 0.644 & 0.644 & 0.631 & 0.622 & 0.613 & 0.605 
& 0.601 & 0.597\\
\hline
$T_{\rm conv}$ (10$^6$ K) \dotfill & 2.20 & 2.19 & 2.19 & 2.20 & 2.20 
& 2.19 & 2.19 & 2.19\\
$P_{\rm conv}$ (10$^{13}$ dyn cm$^{-2}$ ) \dotfill & 5.71 & 5.70 & 5.70 
& 5.69 & 5.68 & 5.68 & 5.67 & 5.66\\
$\rho_{\rm conv}$ (g cm$^{-3}$ ) \dotfill & 0.190 & 0.190 & 0.190 
& 0.189 & 0.189 & 0.189 & 0.189 & 0.189\\
$X_{\rm conv}$ \dotfill & 0.736 & 0.737 & 0.737 & 0.736 & 0.737 & 0.737 
& 0.737 & 0.738\\
$Y_{\rm conv}$ \dotfill & 0.246 & 0.245 & 0.245 & 0.246 & 0.245 & 0.245 
& 0.245 & 0.244\\
$Z_{\rm conv}$ \dotfill & 0.0180 & 0.0180 & 0.0180 & 0.0181 & 0.0181 
& 0.0181 & 0.0181 & 0.0181\\
\hline
Cl(SNU) \dotfill & 7.17 & 6.50 & 5.57 & 4.76 & 3.84 & 3.15 & 2.78 
& 2.45\\
Ga(SNU) \dotfill & 126 & 122 & 117 & 114 & 109 & 105 & 103 & 101\\
$^{8}$B(10$^{6}$ cm$^{-2}$ s$^{-1}$ ) \dotfill & 4.77 & 4.33 & 3.63 
& 3.00 & 2.30 & 1.79 & 1.52 & 1.29\\
   \hline
  \end{tabular}
 \end{center}
$^*$ The first two rows indicate the radius fraction and the mass
fraction of the low-$Z$ core, respectively.
\end{table}

\begin{table}
 \caption{Sensitivity of the neutrino fluxes and the central 
density, $\rho_{\rm c}$, of the low-$Z$
core model with $Z_{\rm c}=0.0001$ and $r_{\rm f}/R_\odot=0.12$ 
($M_{\rm core}/M_\odot=0.117$) to the uncertainties in the input 
physics.$^*$}
 \begin{center}
  \begin{tabular}{rcccccc}
   \hline\hline
  &  &  & Cl    & Ga    & $^8$B & $\rho_{\rm c}$ \\
  &  &  & 3.15  & 105   & 1.79  & 162            \\
Input physics  
  &  &  & (SNU) & (SNU) & (10$^6$cm$^{-2}$s$^{-1}$) 
                                & (g cm$^{-3}$)  \\
\hline
pp           & 4.00(1$^{+0.021} _{-0.013}$)$\times$10$^{-22}$ 
                                 & (keV b) & $_{+0.141} ^{-0.249}$ 
                                                        &
$_{+0.8} ^{-1.5}$ 
           & $_{+0.103} ^{-0.179}$ 
                      & $_{+1.8} ^{-3.4}$ 
                                 \\
pep          & $\pm$1\%          &         & $\pm$0.002 & 
0.0        & 0.000    & 0.0      \\
$^3$He$^3$He & 5.4$\pm$0.32      & (MeV b) & $\mp$0.066 & 
$\mp$0.6 & $\mp$0.040 & $\pm$0.3 \\
$^3$He$^4$He & 0.53$\pm$0.05     & (keV b) & $\pm$0.207 & 
$\pm$1.9 & $\pm$0.125 & $\mp$0.8 \\
$^7$Be+e     & $\pm$2\% (1$\sigma$) 
                                 &         & $\mp$0.040 & 
$\mp$0.1 & $\mp$0.035 & 0.0      \\
$^7$Be+p     & 19$^{+4} _{-2} (1\sigma ) ^{+8}_{-4} (3\sigma )$ 
                                 & (eV b)  & $^{+0.369} _{-0.171}$ 
                                                        & 
$^{+0.8} _{-0.4}$ 
         & $^{+0.347} _{-0.150}$ 
                      & 0.0      \\
$^{12}$C+p   & 1.34$\pm$0.21     & (keV b) & 0.000      & 
0.0      & 0.000      & 0.0      \\
$^{13}$C+p   & 7.6$\pm$1         & (keV b) & 0.000      &
0.0      & 0.000      & 0.0      \\
$^{14}$N+p   & 3.5
$^{+0.4} _{-1.6} (1\sigma ) ^{+1.0} _{-2.0} (3\sigma )$
                                 & (keV b) & $^{+0.000} _{-0.001}$ 
                                                        &  
0.0      & $_{+0.001} ^{-0.000}$ 
                      & 0.0      \\
$^{16}$O+p   & 9.4$\pm$1.7       & (keV b) & 0.000      & 
0.0      & 0.000      & 0.0      \\
$(Z/X)_{\rm surf}$ 
             & 0.0245$\pm$0.0006 &         & $\pm$0.024 & 
$\pm$0.2 & $\pm$0.017 & $\mp$0.2 \\
$L_\odot$    & 3.844(1$\pm$0.004)$\times$10$^{33}$ 
                                 & (erg s$^{-1}$) 
                                           & $\pm$0.082 & 
$\pm$0.8 & $\pm$0.059 & $\pm$0.6 \\
\multicolumn{3}{c}{Sound-speed profile}    & $\pm$0.011 & 
$\pm$0.1 & $\pm$0.008 & $\pm$0.3 \\
$r_{\rm conv} /R_\odot$ 
             & 0.713$\pm$0.001   &         & $\mp$0.005 & 
$\mp$0.1 & $\mp$0.003 & $\pm$0.1      \\
\multicolumn{3}{c}{Neutrino cross-section} 
                                           & $\pm$0.066 & 
$^{+3.1} _{-1.7}$ 
         & $\cdots$   & $\cdots$ \\
Opacity      & $\pm 5\%$         &         & $(\pm)$0.025 
                                                        &
$(\pm)$0.2 
         & $(\pm)$0.018 
                      & $(\pm)$0.1 
                                 \\
EOS          & ideal ($\Gamma = \frac{5}{3}$) 
                                 &         & +0.001     & 
$-$0.2   & $+$0.006   & $-$1.1   \\
\hline
Total        &  $\sqrt{\sigma ^2 + \sigma ^2 + \cdots }$ 
                                 &         & $^{+0.49} _{-0.39}$ 
                                                        & 
$^{+4.0} _{-3.2}$ 
         & $^{+0.39} _{-0.28}$ 
                      & $^{+2.1} _{-3.6}$ 
                                 \\
\multicolumn{2}{c}{(except for opacity and EOS)}  
                                 &         & (SNU)      & 
(SNU)    & (10$^6$cm$^{-2}$s$^{-1}$) 
                      & (g cm$^{-3}$)
                                 \\
   \hline
  \end{tabular}
 \end{center}
$^*$ The first ten entries are due to the nuclear cross-section
 factors. The references for each uncertainty are the same 
as shown in table 4.
\end{table}

\begin{figure}
  \begin{center}
    \FigureFile(40mm,40mm){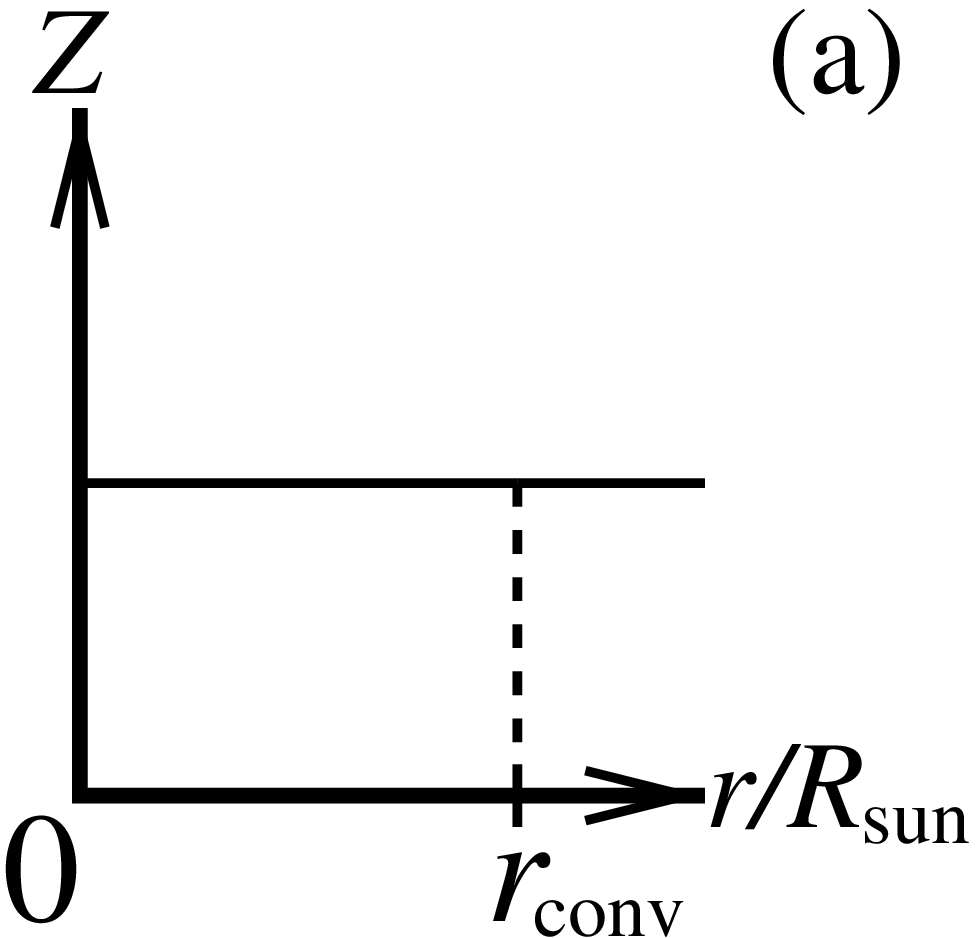}
    \FigureFile(40mm,40mm){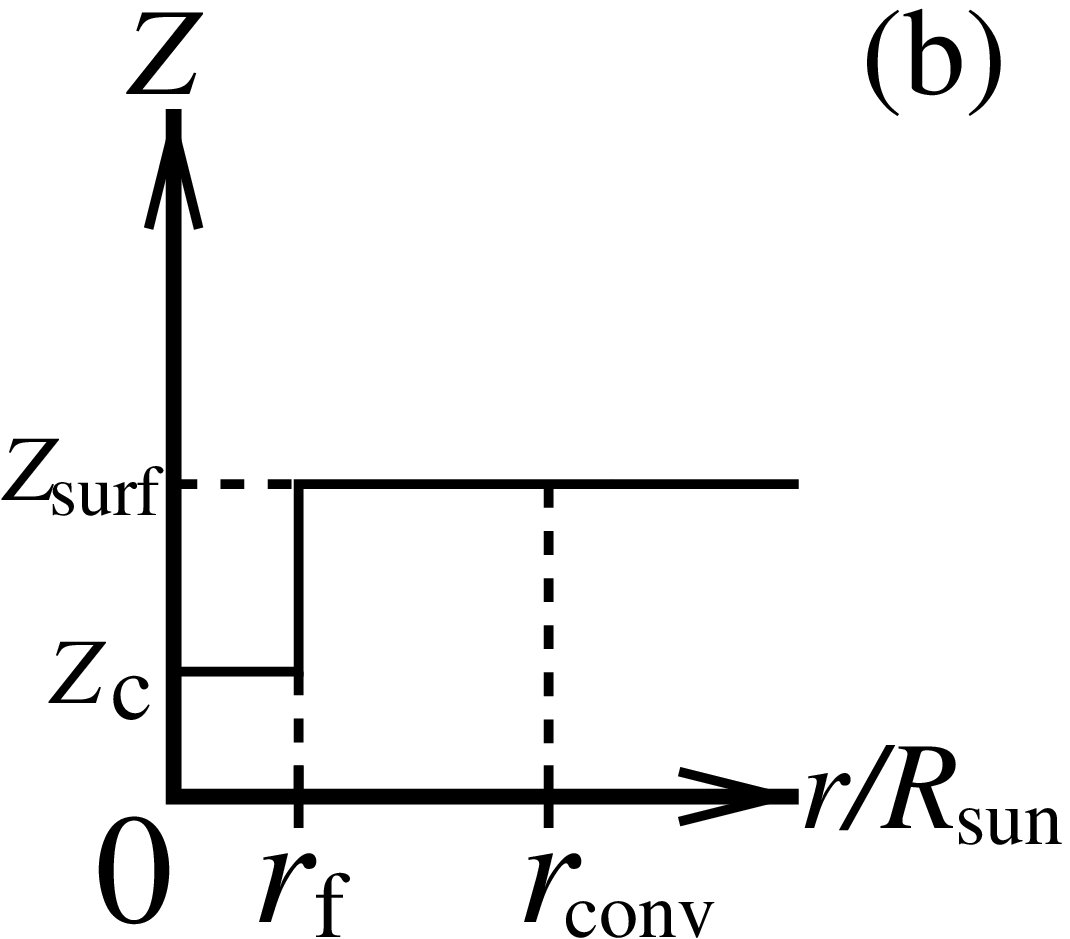}

    \FigureFile(40mm,40mm){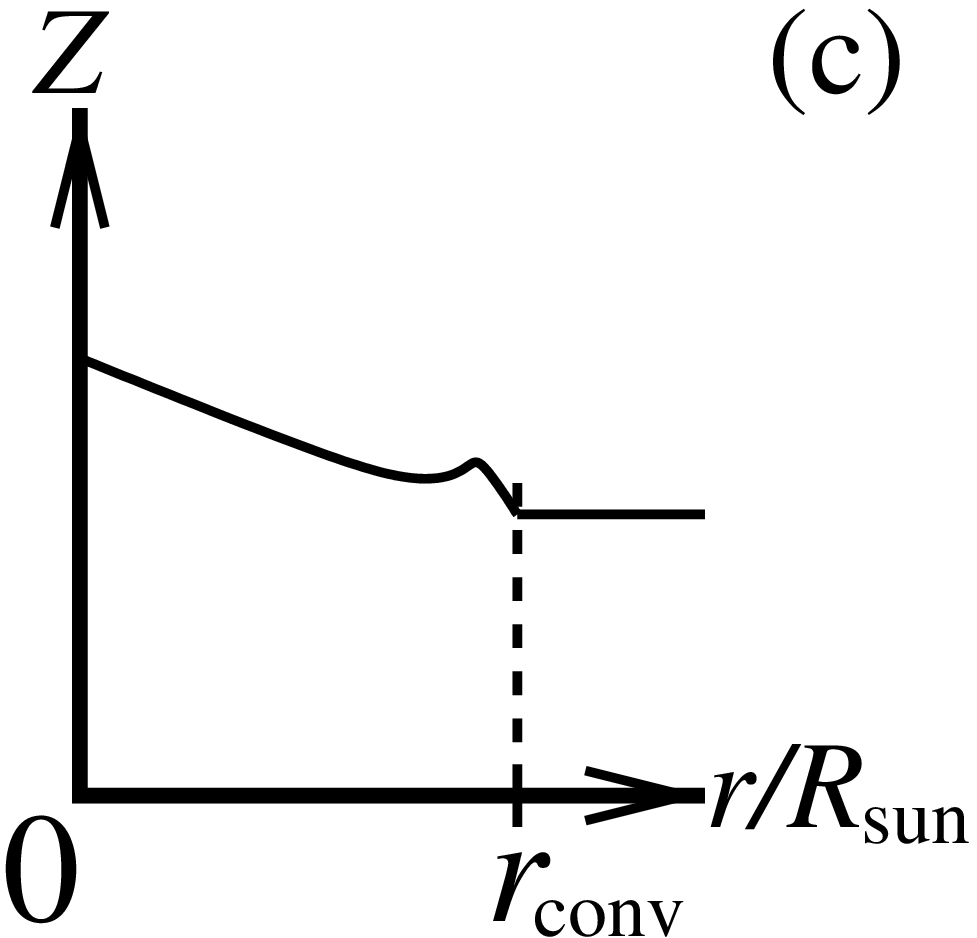}
    \FigureFile(40mm,40mm){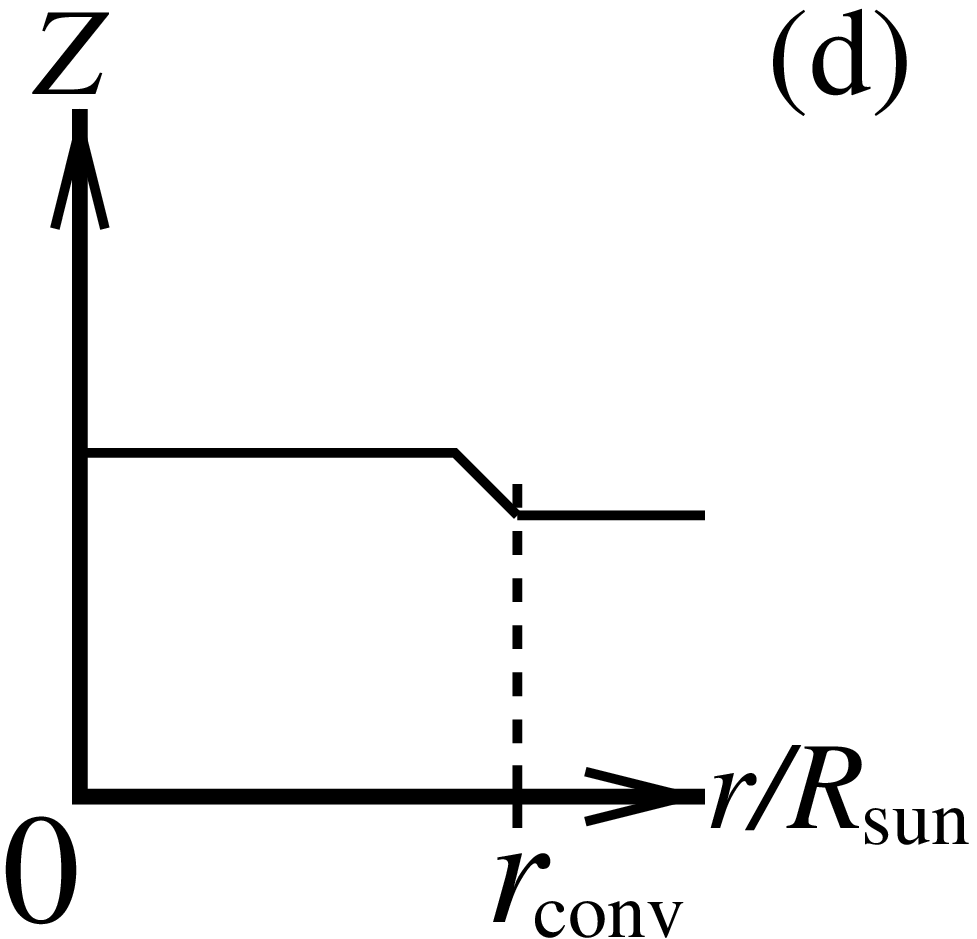}
  \end{center}
  \caption{Comparison of $Z$ distributions.
(a) Homogeneous case (section 3, TS98a). 
(b) Low-$Z$ core (section 4). 
(c) Evolutionary model with gravitational settling (e.g., BP98). 
(d) Rectilinear profile (TS98b).}
\end{figure}

\begin{figure}
  \begin{center}
    \FigureFile(60mm,60mm){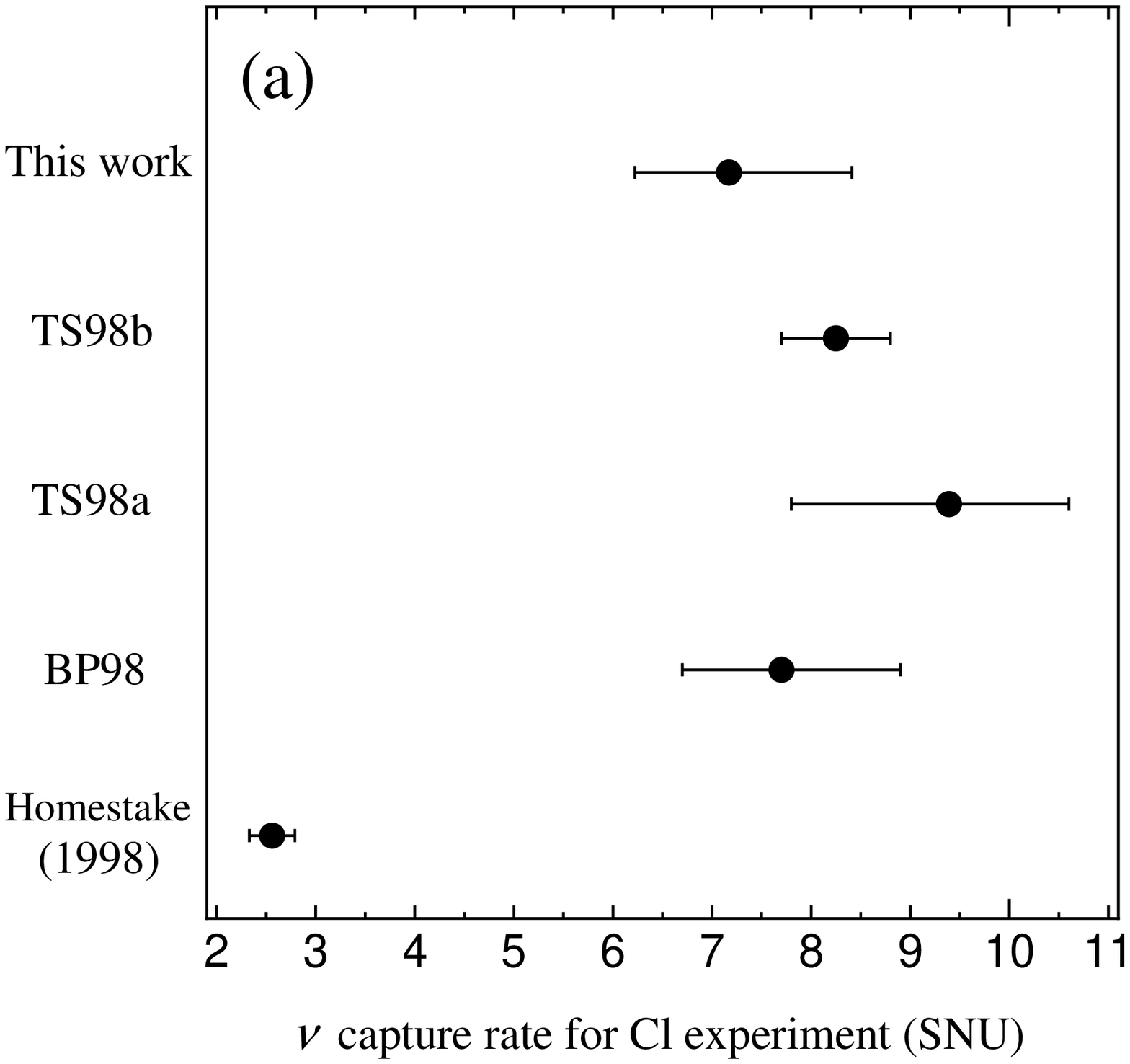}
    \FigureFile(60mm,60mm){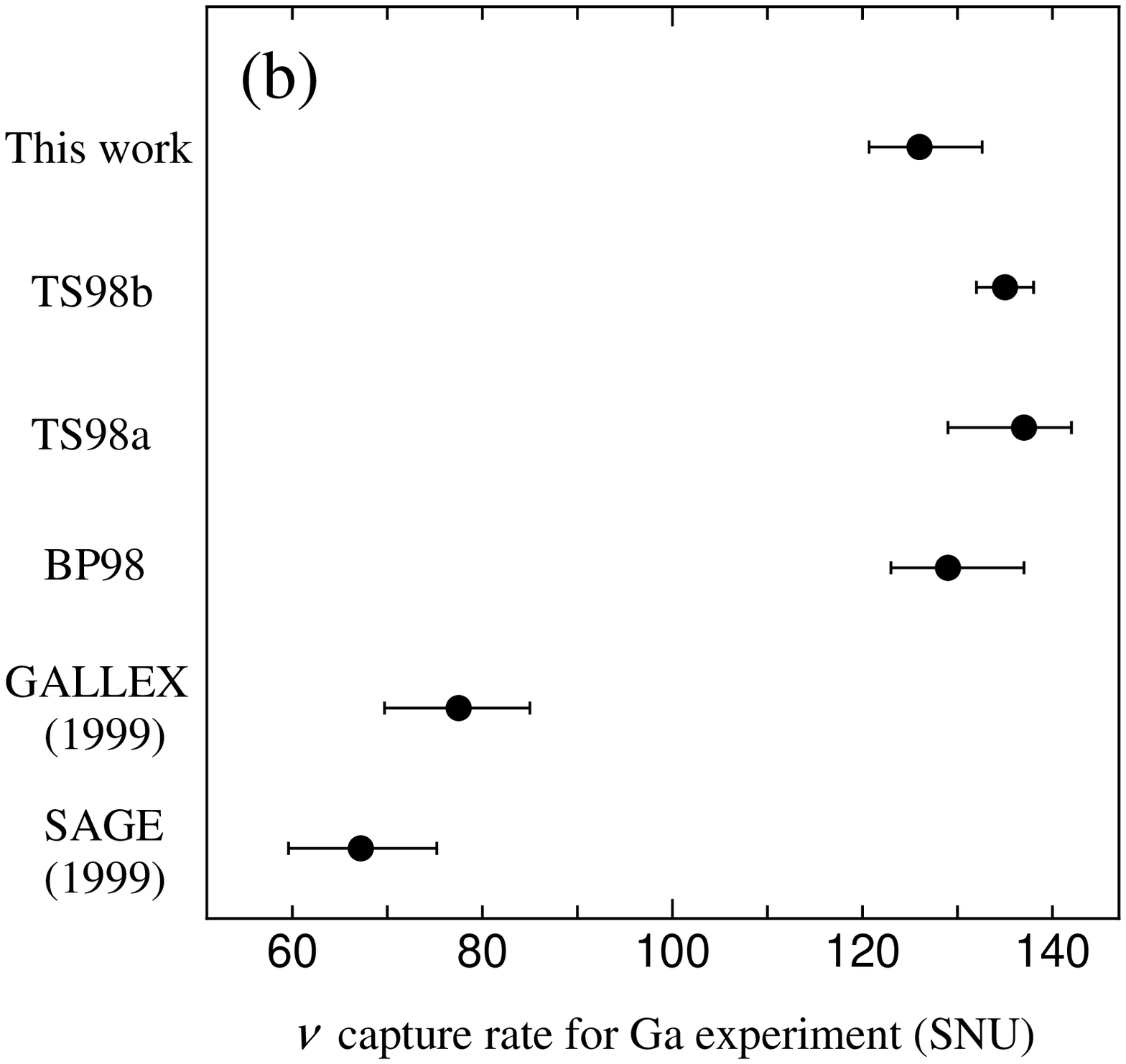}

    \FigureFile(60mm,60mm){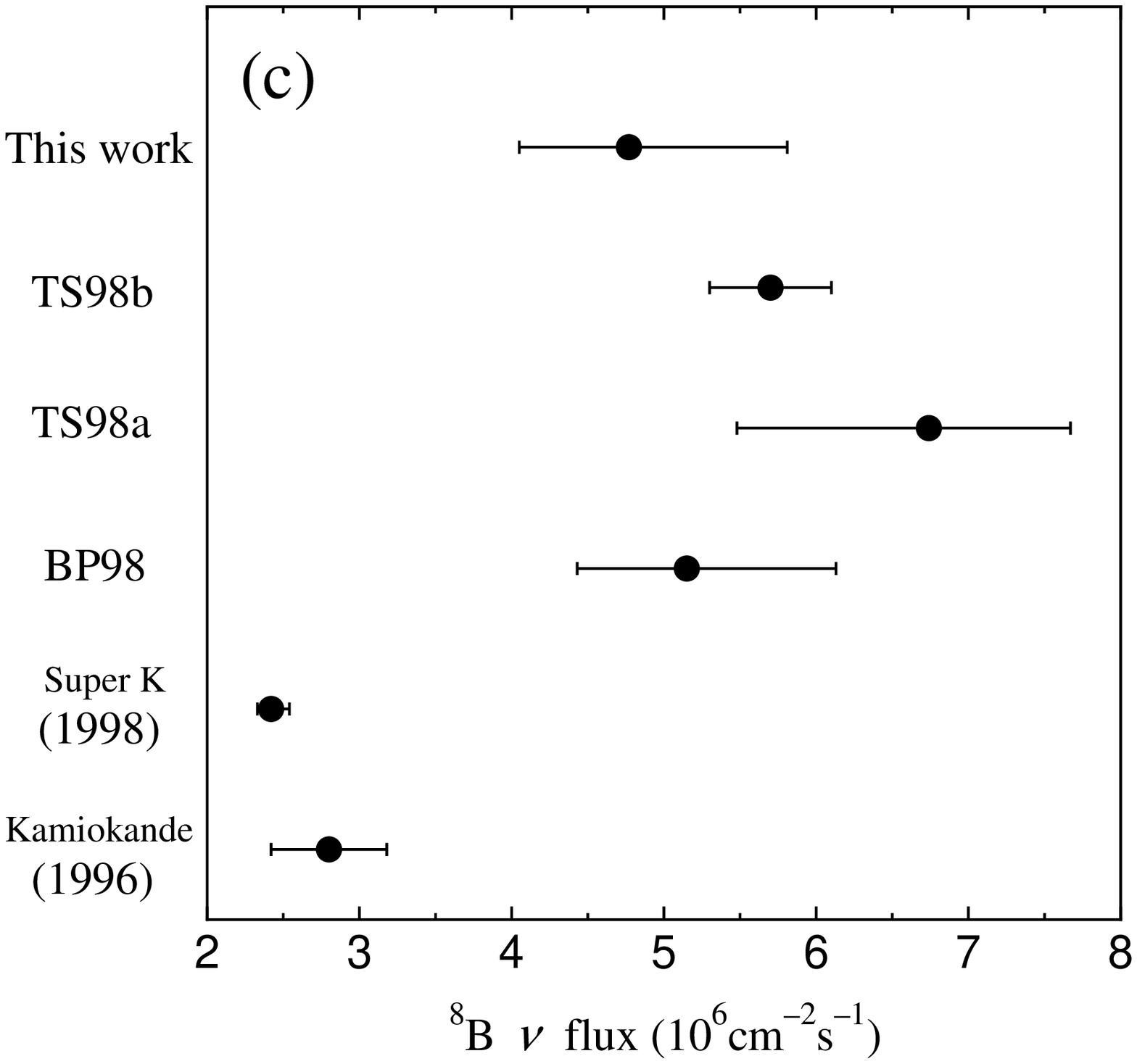}
    \FigureFile(60mm,60mm){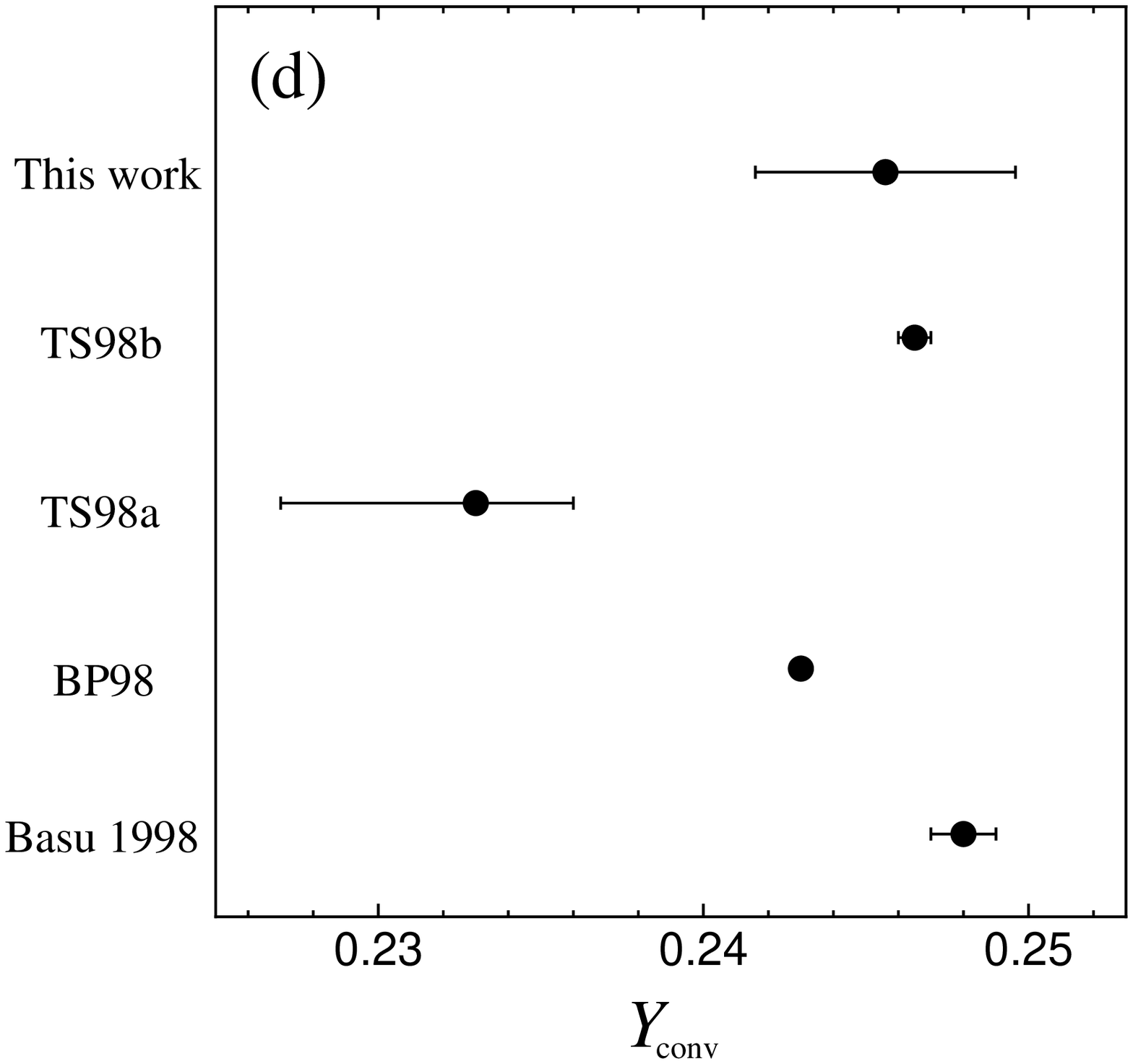}
  \end{center}
  \caption{Comparison of experiments and some solar models.
We should regard the errors of TS98a and TS98b as rough estimations.
(a) Neutrino capture rate for the chlorine detector at Homestake.
(b) Neutrino capture rate for the gallium detectors (GALLEX and 
SAGE).
(c) $^8$B-neutrino flux for the Kamiokande and the Super-Kamiokande.
(d) Helium abundance at the base of the convection zone and the
 surface $Y_{\rm conv}$. For a comparison, the seismically inverted
 abundance, $Y_{\rm conv}$ (Basu 1998), is also shown.}
\end{figure}

\begin{figure}
  \begin{center}
    \FigureFile(65mm,65mm){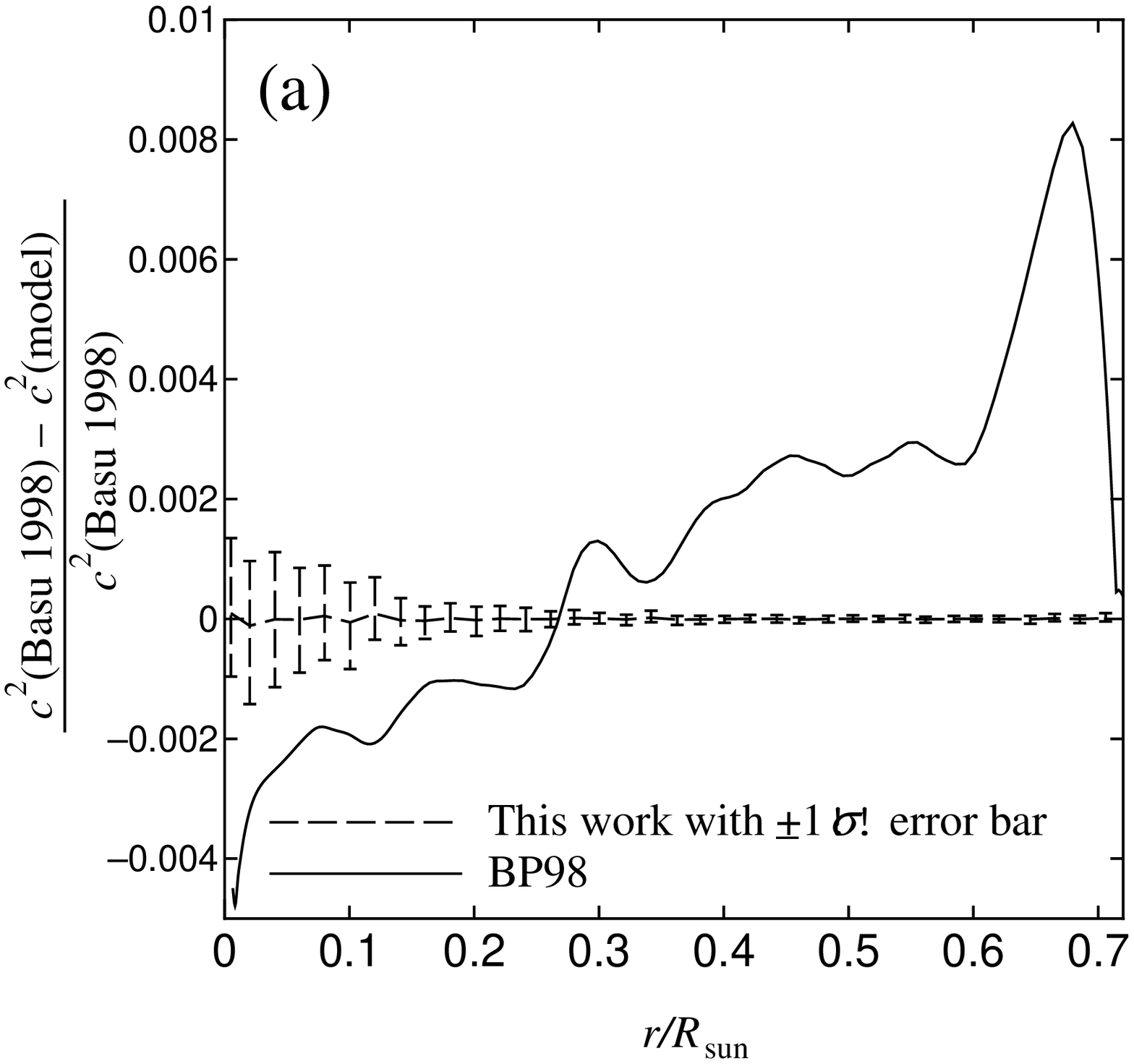}
    \FigureFile(65mm,65mm){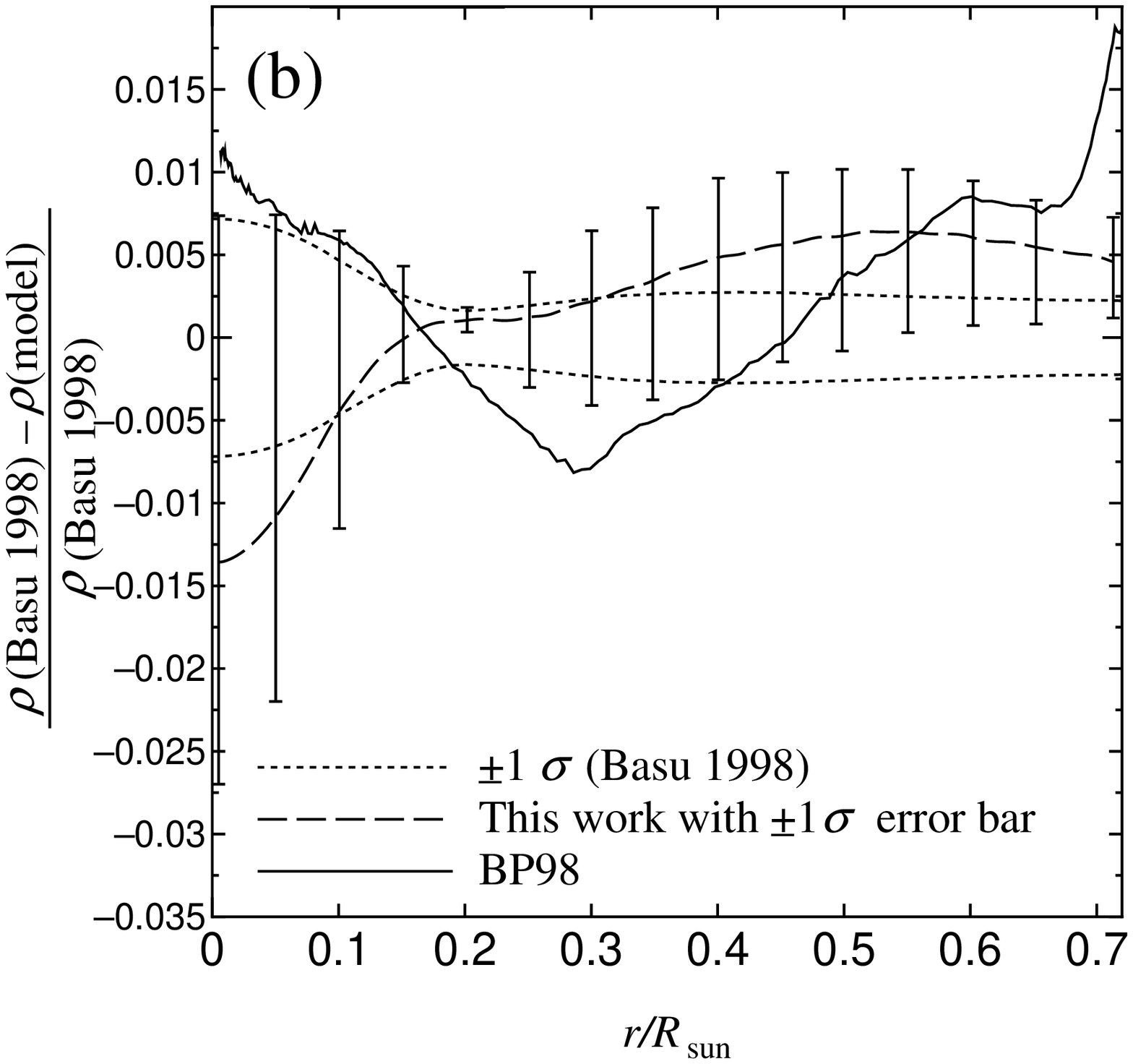}
  \end{center}
  \caption{Relative differences in the squared sound-speed (a) and 
in the density 
(b) between the solar models and the seismically determined 
profiles (Basu 1998). For a comparison, the profiles of an evolutionary 
solar 
model (BP98) are also shown.}
\end{figure}

\begin{figure}
  \begin{center}
    \FigureFile(65mm,65mm){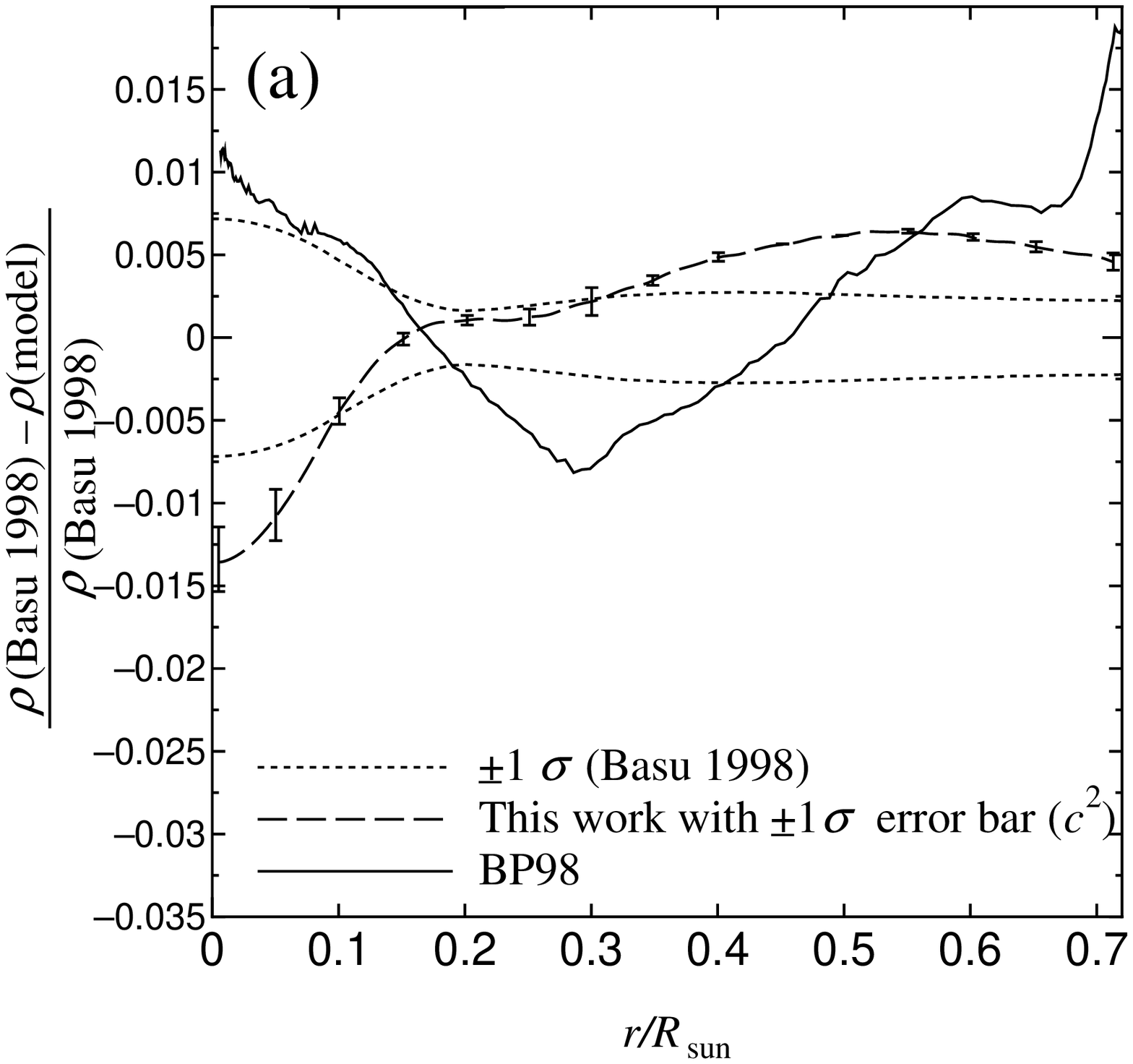}
    \FigureFile(65mm,65mm){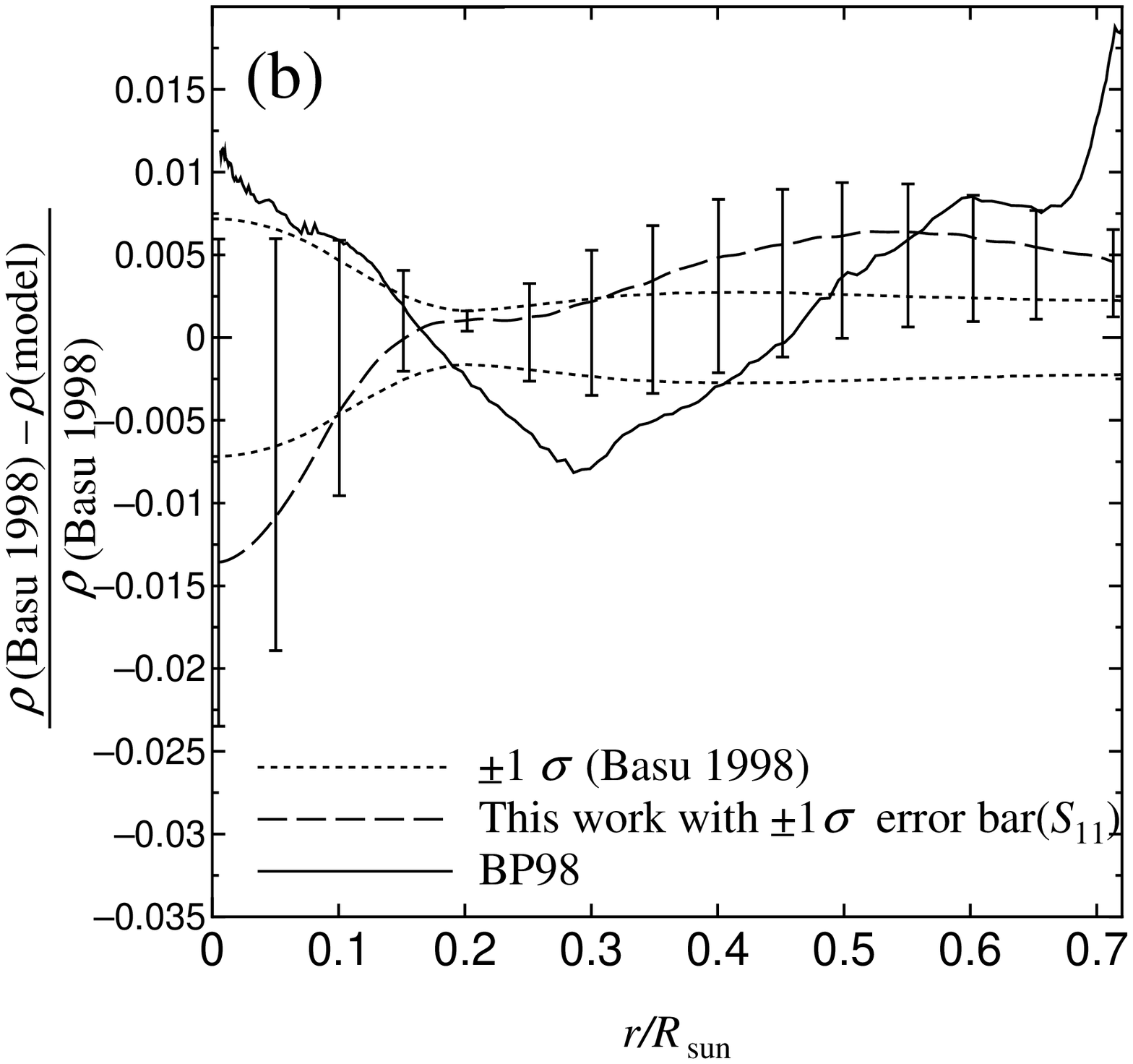}

    \FigureFile(65mm,65mm){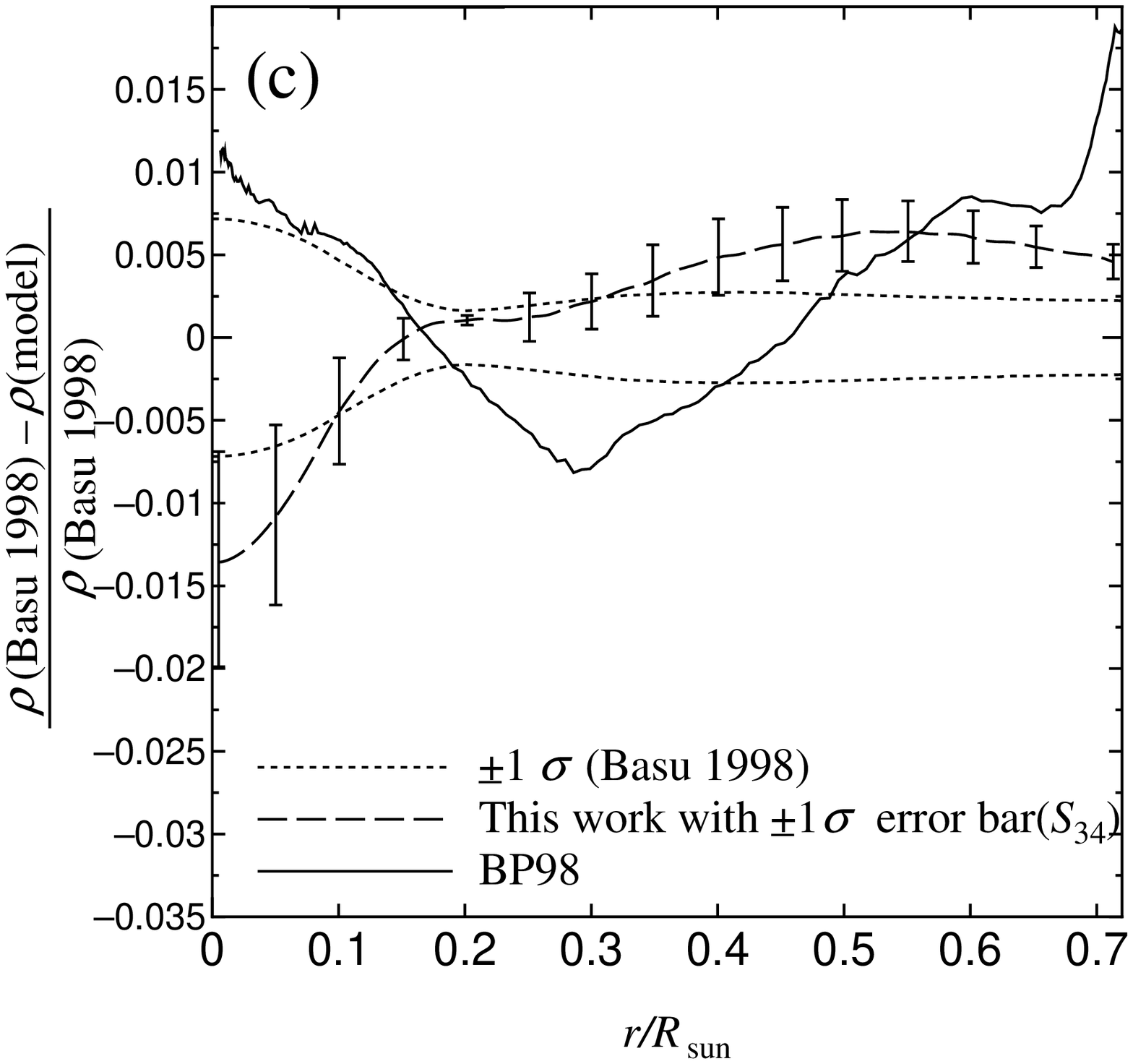}
  \end{center}
  \caption{Uncertainties in the density profile caused only by errors in the 
sound-speed profile (a), by uncertainties in the nuclear 
cross-section $S_{11} (0)$ (b), and $S_{34} (0)$ (c).}
\end{figure}

\begin{figure}
  \begin{center}
    \FigureFile(65mm,75mm){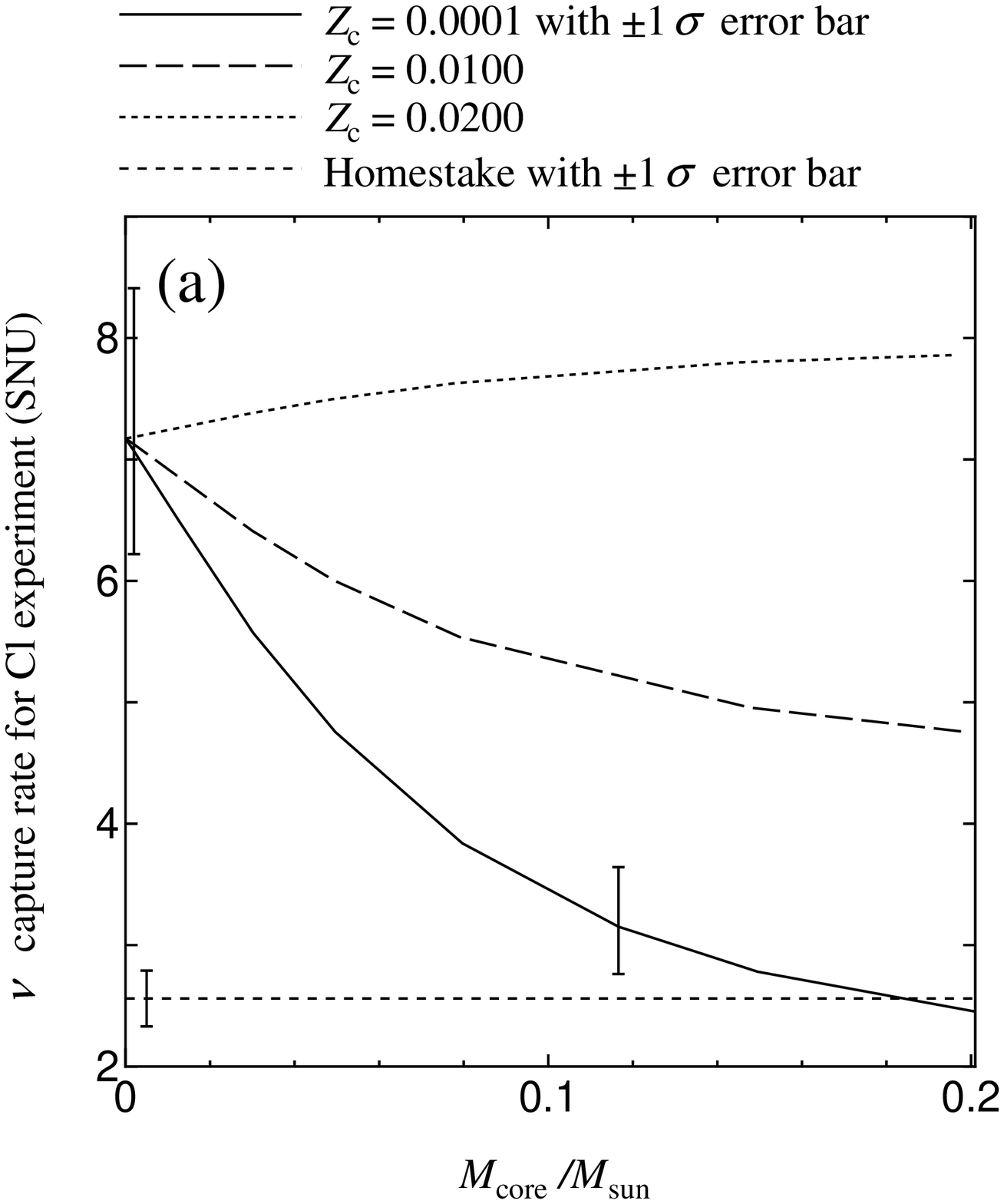}
    \FigureFile(65mm,75mm){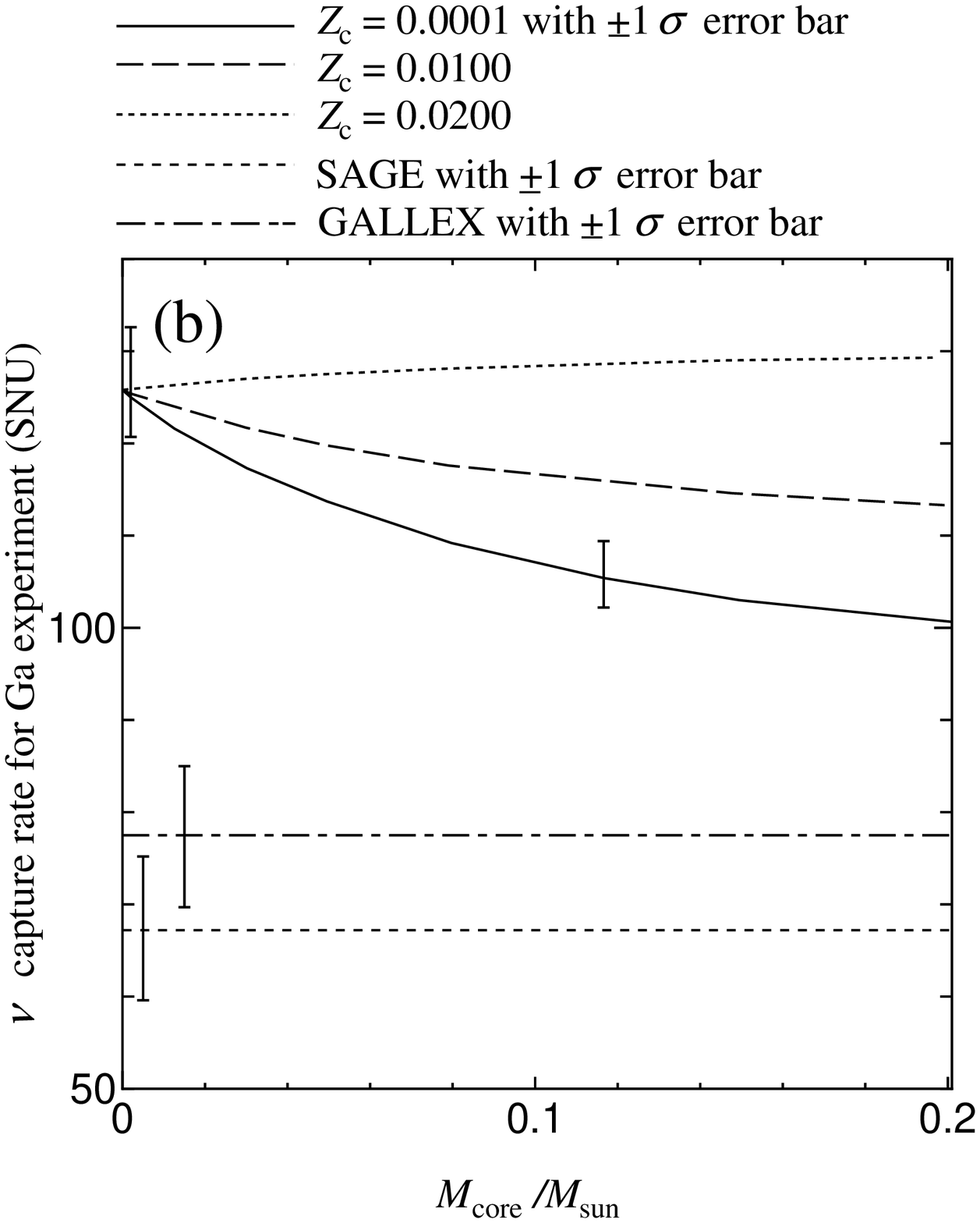}

    \FigureFile(65mm,75mm){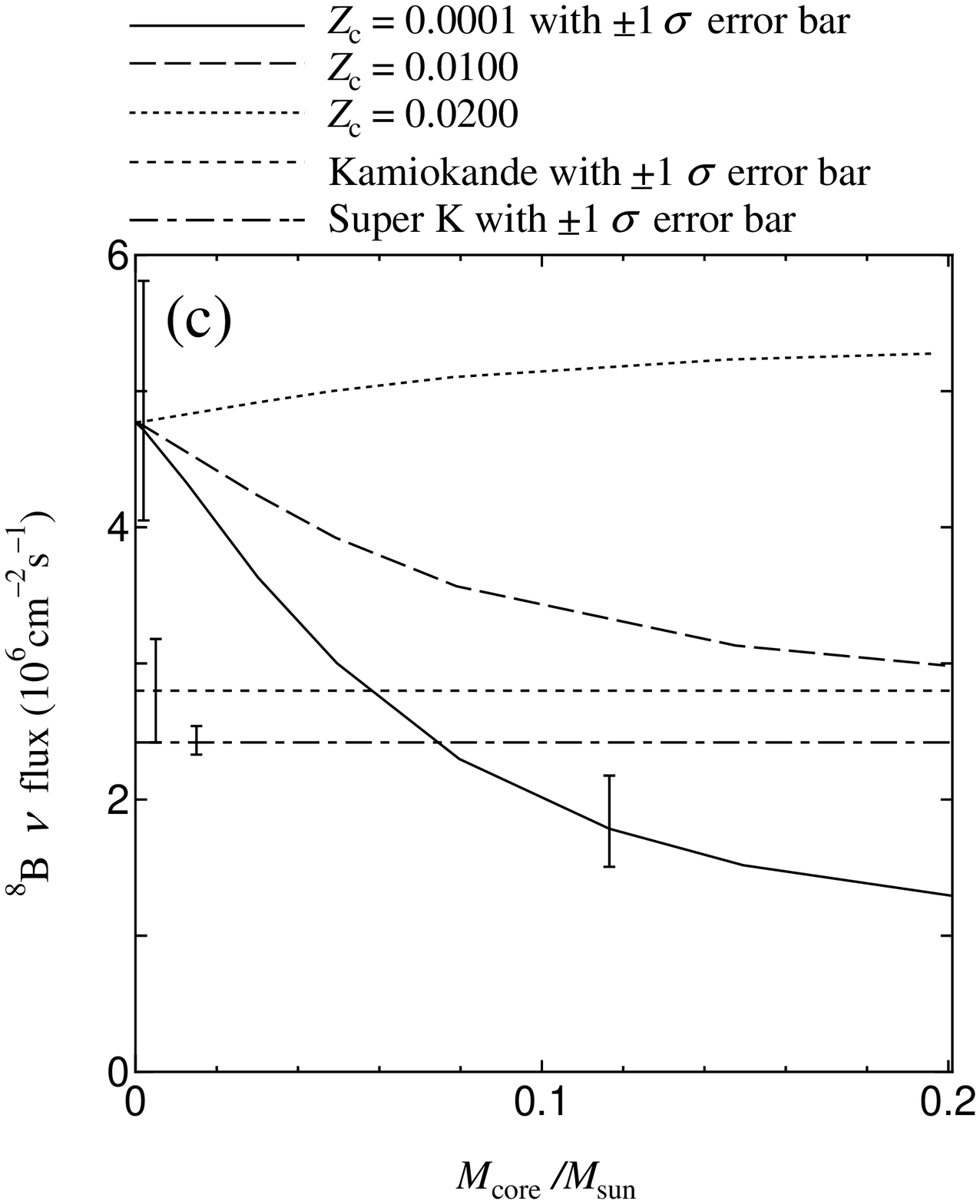}
    \FigureFile(65mm,75mm){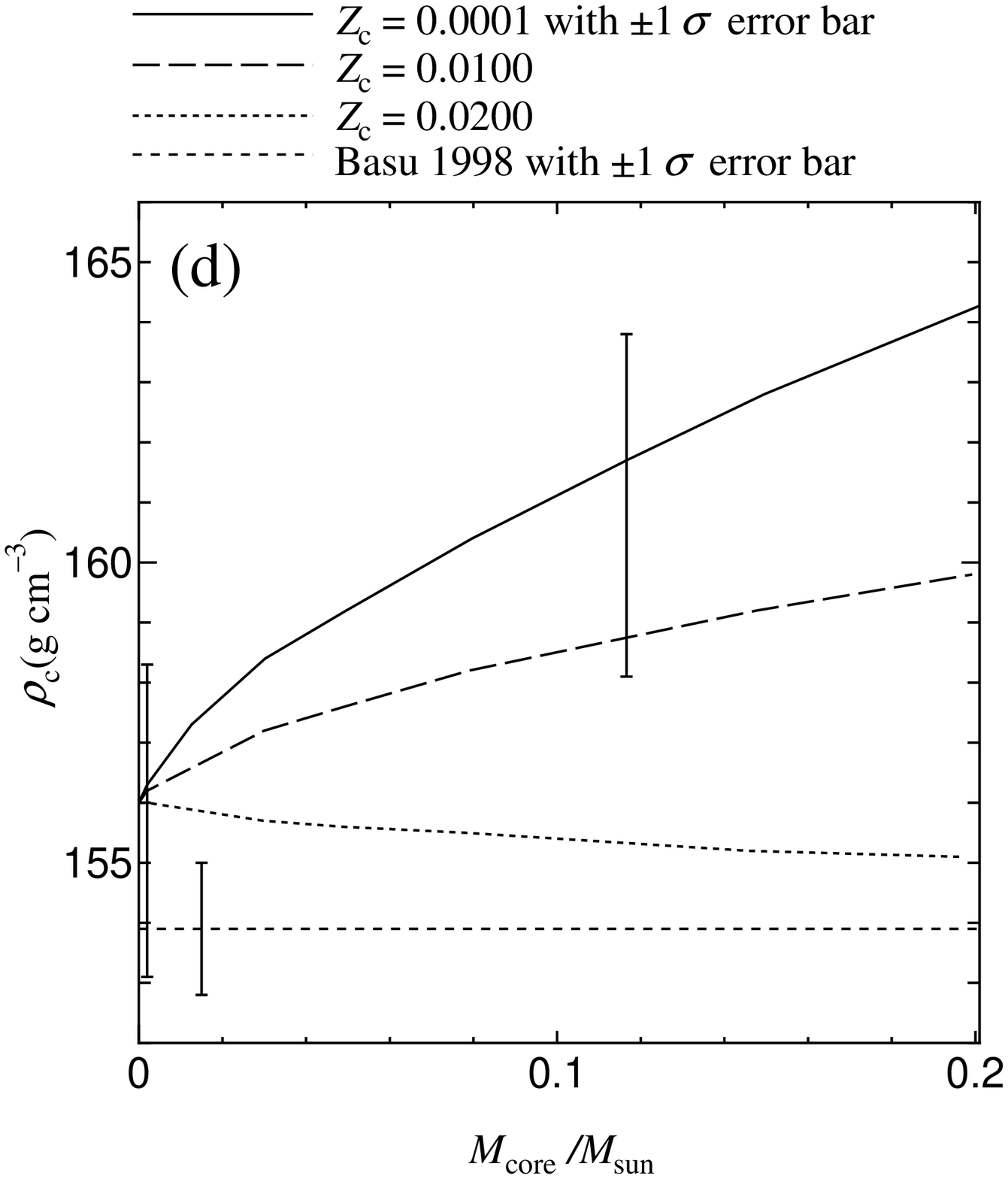}
  \end{center}
  \caption{Dependence of the neutrino fluxes and the central density upon 
$Z_{\rm c}$ and the core size.
(a) Neutrino capture rate for the chlorine detector at Homestake.
(b) Neutrino capture rate for the gallium detectors (GALLEX and 
SAGE).
(c) $^8$B-neutrino flux for the Kamiokande and the Super-Kamiokande.
(d) Central density. For a comparison, the seismically inverted
 density, $\rho _{\rm obs}$
 (Basu 1998), is also shown with the observational error.}
\end{figure}

\end{document}